\documentclass[useAMS,usenatbib]{mnras}
\usepackage{float}

\usepackage[T1]{fontenc}
\usepackage{ae,aecompl}
\usepackage[caption=false]{subfig}

\usepackage{graphicx}	
\usepackage{amsmath}	
\usepackage{amssymb}	
\usepackage{natbib}
\usepackage{float}
\stepcounter{equation}

\newcommand{\Rip}{$R_{\rm ip}$}
\newcommand{\Rbg}{$R_{\rm 3\sigma}$}
\newcommand{\Rfit}{$R_{\rm fit}$}
\newcommand{\Ibg}{$I_{\rm bg}$}

\newcommand{\sigmabg}{$\sigma_{\rm bg}$}

\title[Structural parameters of M82 disk SSCs]
{Structural analysis of disk super star clusters of M82: size and profile shape at intermediate ages}

\author[Cuevas-Otahola et al.]{
B. Cuevas-Otahola,$^{1}$\thanks{E-mail: bolivia@inaoep.mx}
Y. D. Mayya,$^{1}$
I. Puerari$^{1}$
and D. Rosa-Gonz\'alez$^{1}$
\\
$^{1}$Instituto Nacional de Astrof\'isica, \'Optica y Electr\'onica, 72840 Puebla, Mexico\\
}

\date{Accepted XXX. Received YYY; in original form ZZZ}

\pubyear{2019}

\begin{document}
\label{firstpage}
\pagerange{\pageref{firstpage}--\pageref{lastpage}}
\maketitle

\begin{abstract}
We present the structural parameters of 99 Super Star Clusters (SSCs) in the
Disk of M82. Moffat-EFF, King and Wilson models were fitted using a $\chi^2$
minimisation method to background-subtracted Surface Brightness Profiles (SBPs)
in F435W (B), F555W (V), and F814W (I) bands of the Advanced Camera for Surveys
(ACS) of the Hubble Space Telescope (HST).  The majority of the SSC profiles is
best-fitted by the Moffat-EFF profile.
The scale parameter $r_{\rm d}$ and the shape parameter $\gamma$ in the three
filters are identical within the measurement errors.
The analysed sample is big enough to allow characterisation of the
distributions of
core radii $R_{\rm c}$ and $\gamma$.
The obtained distribution of $R_{\rm c}$ follows a log-normal form,
with center and $\sigma log\big{(}\frac{R_{\rm c}}{pc}\big{)}$ being 1.73~pc
and 0.25, respectively. The $\gamma$ distribution is also log-normal with center
and $\sigma log(\gamma)$ being 2.88 and 0.08, respectively.
M82 is well-known for the absence of current star formation in its
disk, with all disk
SSCs older than 50~Myr and hardly any cluster older than $\sim$300~Myr.
The derived distributions compare very well with the distributions for intermediate-age clusters
in the Large Magellanic Cloud (LMC), which is also a low-mass late-type galaxy
similar to M82.
On the other hand, the distributions of $R_{\rm c}$ in both these
galaxies are shifted
towards larger values as compared to SSCs of similar age  in the giant spiral galaxy
M83. M82 and LMC also
span a narrower range of $\gamma$ values as compared to that in M83.

\end{abstract}




\section{Introduction}

Understanding the formation and evolution of globular clusters (GCs) has been 
an active field of research in astrophysics over the last half century 
\citep{ForbesGCreview2018}. The discovery of clusters as dense 
($\rho \gtrsim10^3 \rm M_\odot /pc^3$) and massive ($10^4-10^6 \rm M_\odot$)
as GCs, but relatively young, known as Young Massive Clusters or 
Super Star Clusters (SSCs), has given a new impetus to these studies in the 
last two decades \citep{Bastianrev2016}.
SSCs are often thought to be the progenitors of GCs, and 
hence their study has the potential to throw light on the processes that the 
GCs may have experienced during their early evolution \citep{Portegiesrev}.
SSCs are subjected to  different physical processes at different timescales: 
at short ($\lesssim 10^7$~yr) and intermediate timescales ($10^7$--$10^8$ yr), 
stellar evolutionary processes (stellar winds, supernovae, etc) play a role; 
at later times ($\gtrsim10^8$~yr) dynamical processes start becoming important: 
the most dominant of them being the gravitational shocks due to the 
interaction of the cluster with the tidal field of its host galaxy, and two-body 
relaxation \citep{Spitzer_book_gcs}. These processes increase the velocity of some stars above the escape 
velocity, forcing them to leave the cluster, resulting in the dissipation and/or 
complete disruption of the cluster. The selective loss of 
high-velocity stars from the central regions of the clusters leads to collapse
of the core at late times \citep{Lynden-Bell1968}.
The extent to which a cluster is subjected 
to these effects depends on its 3-dimensional location in 
its host galaxy, galacto-centric distance, in addition to the gravitational potential field of the host
galaxy itself \citep{Fall&Zhang}. Besides, clusters located in the disk 
suffer from encounters with Giant Molecular Clouds when they pass through the
spiral arms \citep{Gieles2006}. \cite{Mackey2008} analysed the effect 
of binary and single black holes and found them to be responsible of expansion
of the core at times $\gtrsim$ 600~Myr in clusters in the Large Magellanic Clouds
(LMC).   In the presence of a tidal field, the sizes of the expanding clusters would be limited to their tidal radius \citep{Gieles2013}.

The structure of star clusters has been modelled theoretically using 
auto-gravitating isothermal spheres of lowered kinetic energy in the
presence of external tidal forces. These configurations, usually known
as King models following the classical treatment of \cite{King_dyn},
explain satisfactorily the observed surface brightness
profiles (SBPs) of old stellar systems such as GCs \citep{Baumgardt2018}. 
The most salient
feature of these SBPs is the existence of a core-halo structure, with 
the core characterised by the core radius and the halo limited by
the tidal radius. On the other hand, SBPs of slightly younger systems such 
as the blue population of clusters in the LMC, lack core-halo structure and instead follow
power-law forms. \citet{Elson} found that these SBPs are 
well-represented by Moffat profiles (Moffat-EFF profiles, henceforth).
The profile of the most massive and luminous SSC in the LMC, R136, is
also consistent with a Moffat-EFF profile \citep{Elson1992}.
\cite{MackeyGilmore2003a} fitted Moffat-EFF profiles to the
SBPs of 53 star clusters in the LMC to obtain their structural parameters.
Wilson profiles, originally proposed
by \citet{Wilson_dyn} to characterise SBPs of elliptical galaxies, are also
found to be good fits to the SBPs of SSCs in the LMC \citep{McLaughlin2005}.
The power-law nature of SBPs at relatively younger ages is understood to
be due to the contribution of stars in the unbound halo \citep{Elson, MorenoPichardo2014}.

Hubble Space Telescope (HST), especially the wide field of the
Advanced Camera for Surveys (ACS), has enabled
the detection of large populations of SSCs in external galaxies, some examples
being M82 \citep{Oconnell1995,Melo2005,Mayyacat}, M51 \citep{Chandar2011}, 
M81 \citep{Chandar2001,SantiagoCortes2010}, 
M83 \citep{Bastian2011,Ryon},  and Antennae (NGC4038/4039) \citep{Whitmore&Schweizer1995}.
Clusters have been reported in 20 other nearby spiral and irregular galaxies
using the Hubble Legacy Archive (HLA) data \citep{Whitmore2016}.

Modern $\chi^2$ minimisation technique allows the analysis of the SBPs of SSCs with
empirical formulae in an objective way. Moffat-EFF and empirical formulae
for King models \citep{King_emp}, available directly in profile analysing 
tools such as ISHAPE \citep{Larsenishape} and GALFIT \citep{Penggalfit},
are the most often used profiles 
for fitting SBPs of SSCs. The output parameters commonly obtained
by such analysis are core radius and half-light radius,
often for an assumed form of the profile shape \citep[e.g.][]{BastianGieles2008}. 
\citet{Ryon} and \citet{Ryon2017} carried out the analysis of structural parameters 
on the HST images of $\sim$700 Young Massive Clusters (YMCs) in the giant spiral M83, and in 
two late-type galaxies (NGC~628 and NGC~1313). 
They obtained core radius, half-light radius, and the shape parameters 
using Moffat-EFF profile in GALFIT for 478 YMCs that are well resolved on the HST images. 
For the rest of the YMCs, they obtained half-light radius based on 
an empirical relation between the concentration index, defined as the
difference in magnitudes between 1 and 3 pixel radius apertures, and
half-light radius on mock YMCs. YMCs they analysed are in general younger than 1~Gyr, and constitute the largest sample of intermediate-age YMCs with uniformly
determined structural parameters.

\cite{McLaughlin2005} proposed an alternative technique to 
obtain structural parameters of star clusters. Their method involves fitting
the observed profiles directly
with the profiles generated using dynamical models that have underlying physical
basis such as \cite{King_dyn} and \cite{Wilson_dyn}. They also suggested using Jeans theorem
to construct dynamical models that are consistent with the empirical Moffat-EFF profile.
They used this technique to obtain a complete set of dynamical parameters, not just
core radius and half-light radius, for GCs in the Milky Way and Fornax galaxies, and blue and
red star clusters in the SMC and LMC galaxies.
The technique has been recently extended by \citet{Sollima2015} to implement 
anisotropic King-Michie models. 
For clusters of known age, and hence known photometric masses, this technique 
is able to extract the central and line-of-sight velocity dispersions. The latter
parameter could be determined observationally using high spectral resolution
observations, which allows a direct verification of the validity of the assumption
of the dynamical models used. 
The HST images of galaxies that are nearer than $\sim$5~Mpc have sufficient 
spatial resolution (1~pix=1.25~pc at 5~Mpc) for the construction of SBPs good enough not only for the 
determination of their sizes, but also for a detailed analysis using dynamical models.
Beyond the Milky Way, M31 and NGC5128 are the only two giant galaxies where SSC 
profiles have been analysed using dynamical models \citep{Barmby2007,Mclaughlin2008,Wang_2013}.

M82 is an excellent candidate to carry out such a study, as it is 
relatively nearby \citep[3.63~Mpc][]{Freedman}, and has a rich population of 
clusters in its nucleus and disk \citep{Mayyacat}.  
Spectroscopic ages have been obtained for around 40 of the disk SSCs.
The derived ages occupy a relatively narrow range
between 50--300~Myr \citep{Konst2009}. \citet{Mayya2006} suggested, based on the 
analysis of the photometric, dynamical, and chemical properties, that
the entire galaxy participated in a disk-wide burst of duration of a few hundred
million years. The disk stopped forming stars around 50~Myr ago, which is
well supported by the absence of red supergiants in its disk \citep{Davidge2008}. 
The disk-wide burst, and the formation of disk clusters, were most likely triggered  
by the interaction of M82 with its neighbours M\,81 and NGC\,3077 
\citep{Yun1999}. The narrow age range of disk SSCs is consistent with
them being formed in the disk-wide burst.
The existence of a few hundreds of massive
SSCs, all of ages intermediate between the YMCs and the old GCs, gives us a great 
opportunity to understand the dynamical effects experienced by evolving clusters.

In \S\ref{sec:sample_gen}, we summarise the general properties of the sample 
of SSCs in the disk of M82, as well as the procedure followed in this work
to obtain the background-subtracted SBPs. 
In \S\ref{Sec:method_fit}, we describe the SBPs expected in theoretical models such as King and Wilson,
and empirical formulae (Moffat-EFF), as well as the procedure followed to obtain the
structural parameters from the observed SBPs.   
Model-derived parameters are 
presented and their statistical properties discussed  in \S\ref{Sec:results}. The results are summarised in \S\ref{Sec:summary}.
    
\section{M82 SSC sample and extraction of surface brightness profiles}\label{sec:sample_gen}

The sample of SSCs for structural analysis was selected from the catalogue of M82 disk 
SSCs from \cite{Mayyacat}, which consists of 393 objects, and is based 
on the detection of SSCs in F435W (B), F555W (V) and F814W (I) bands of the HST/ACS.  The entire sample is presented in Table 3 of \citet{Mayyacat}.   In \S\ref{Sec:gen_char}, we will compare the magnitude and colour properties of the subsample with respect to the entire sample.

\subsection{Surface brightness profiles}\label{Sec:sbps}

We used the same images which were used for cluster detection to extract SBPs.   These images were part of the HST Legacy Survey, which were made available 
in reduced form by the Hubble Heritage Team \citep{Mutchler2007}.  The 
image scale corresponds to 0.05\arcsec\,pixel$^{-1}$ and covers the 
entire optical extent of M82.  Exposure times were 1600, 1360 and 1360 seconds in 
filters B, V and I, respectively.  The zero-point magnitudes in the Vega system 
were extracted from \cite{Sirianni2005}, with values 25.779, 25.724, and 25.501 
for B, V, and I bands, respectively.  

\begin{figure*}
\begin{center}
\subfloat{\includegraphics[width=0.55\textwidth]{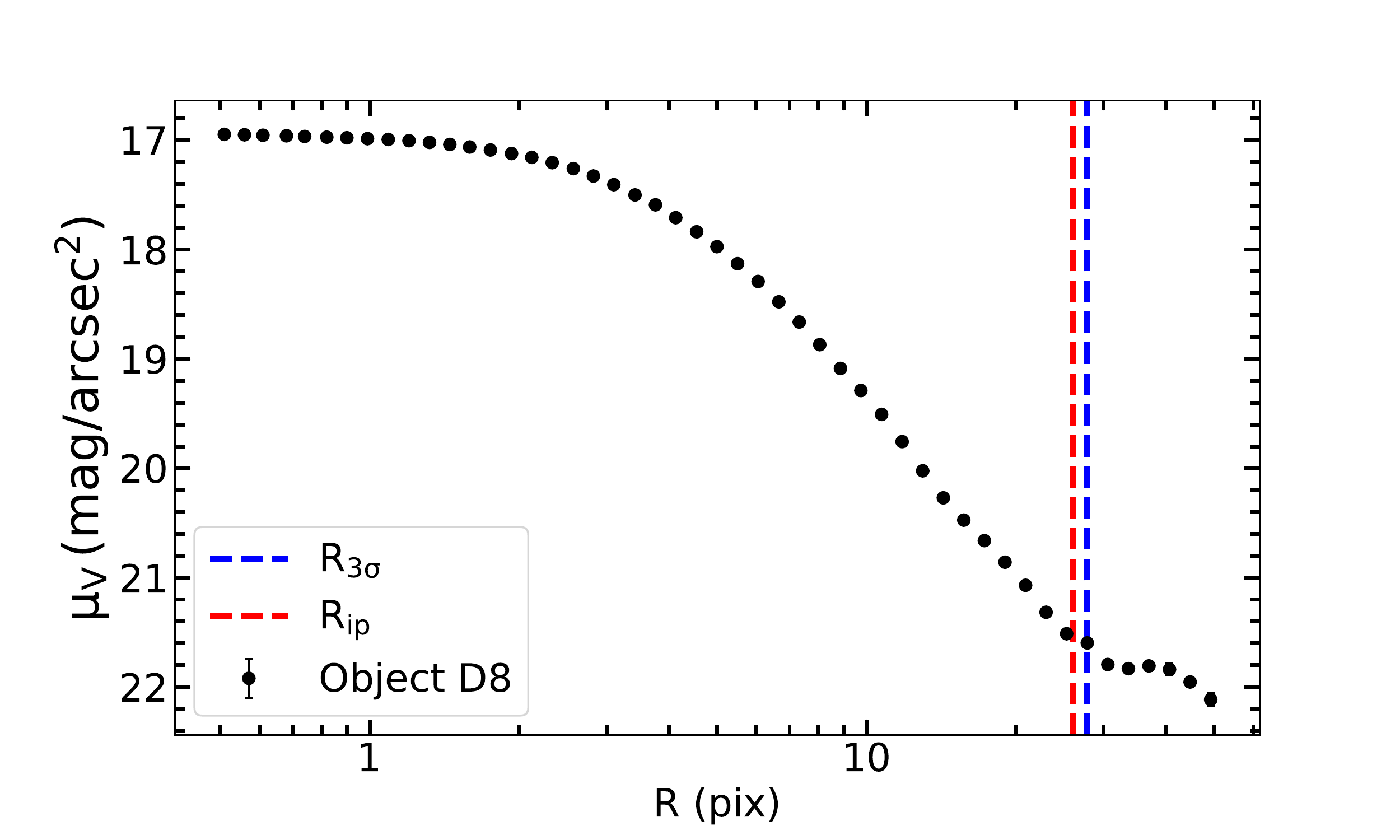}}
\label{bla1}
\subfloat{\includegraphics[width=0.3\textwidth]{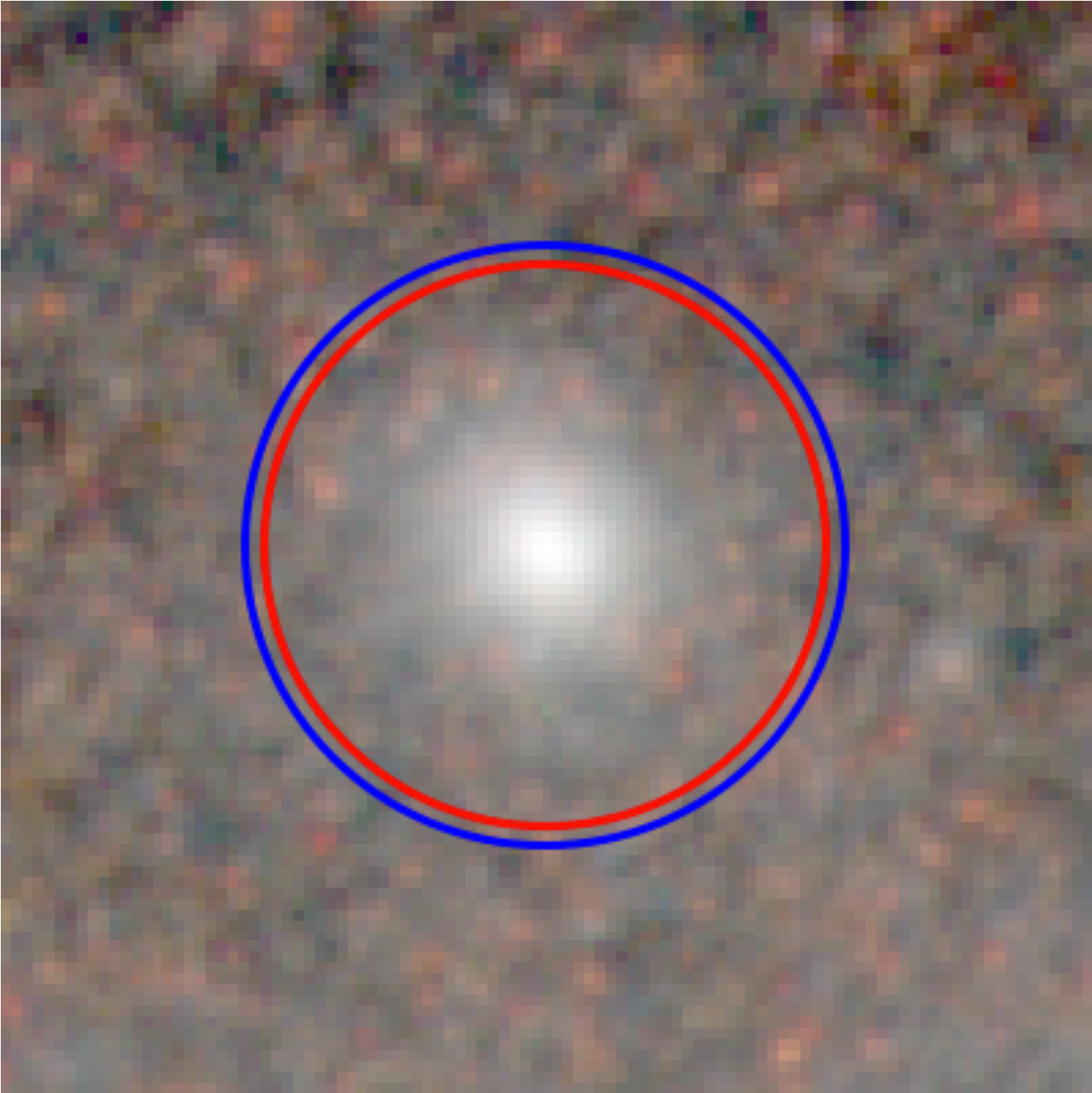}}\label{bla2}
\subfloat{\includegraphics[width=0.55\textwidth]{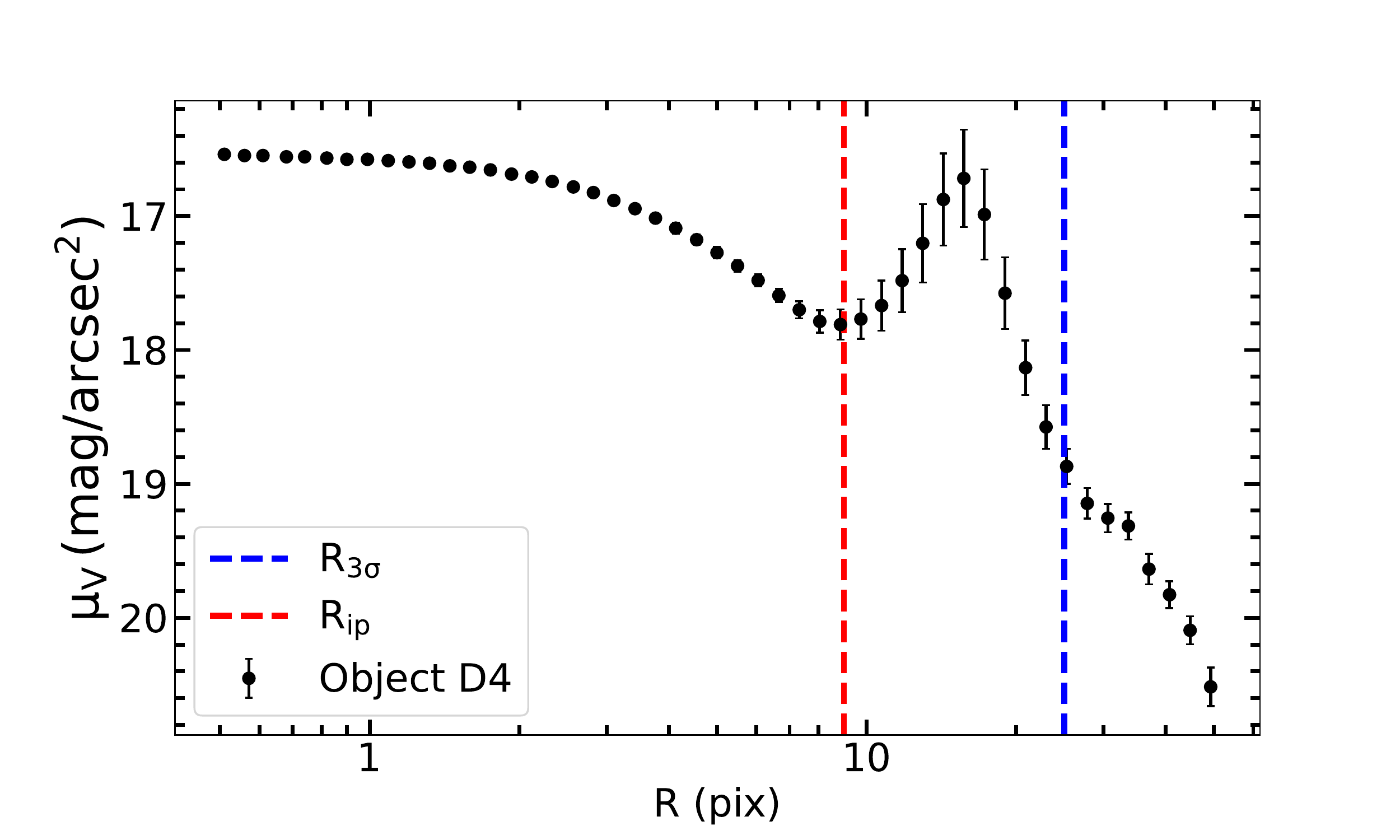}}
\label{bla3}
\subfloat{\includegraphics[width=0.3\textwidth]{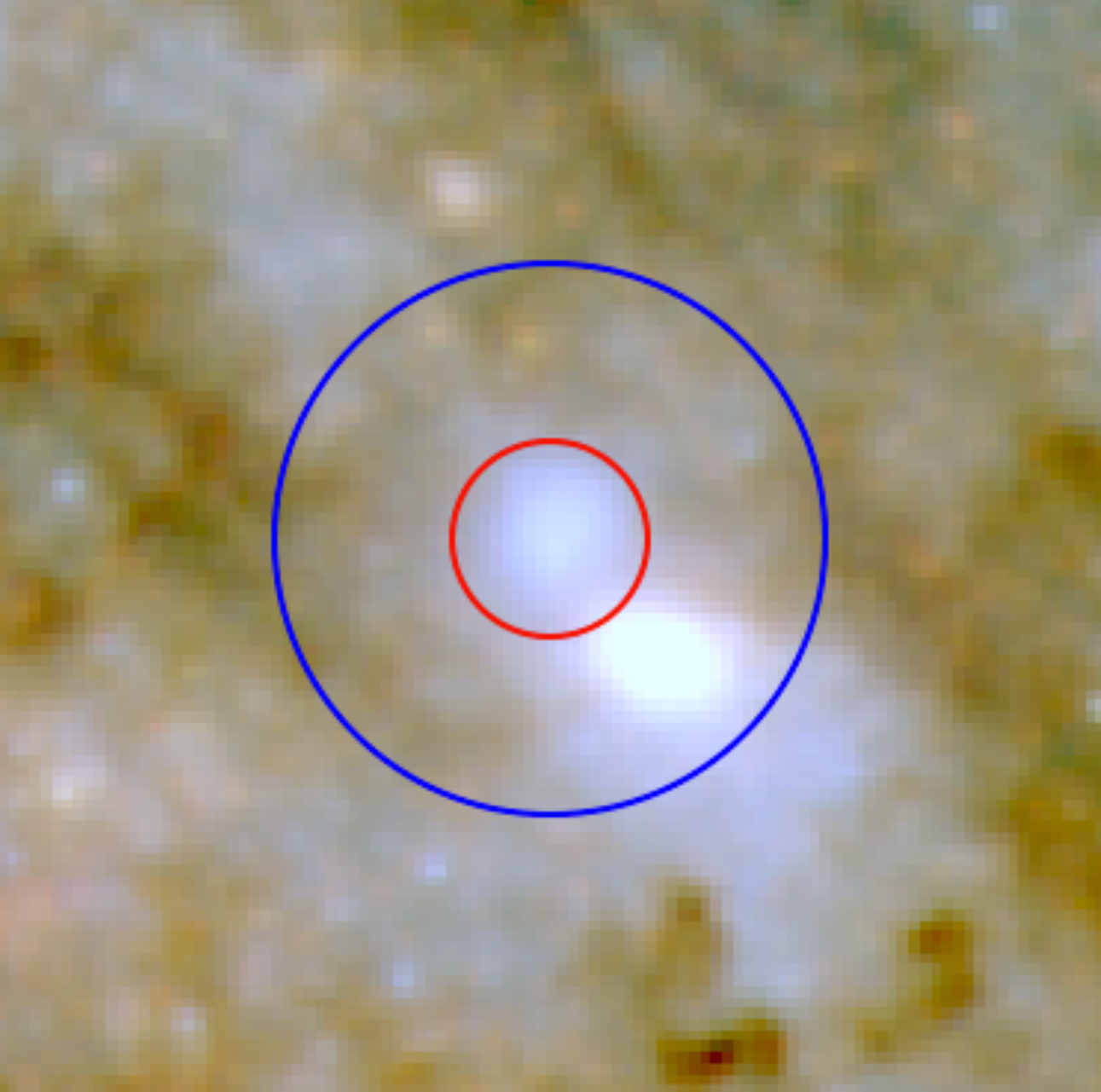}}\label{bla4}
\caption{The $V$-band surface brightness profiles (left) and RGB image, constructed from I (R), V (G) and B (B) bands  (right) of two clusters, illustrating the procedure adopted for selecting the fitting radius. In the top,
we illustrate it for the SSC D8, an isolated cluster, whereas in the
bottom, we illustrate
it for a highly-contaminated, but still useful, SSC D4. The
$R_{3\sigma}$, the radius
where the background-subtracted intensity is 3$\times \sigma_{\rm bg}$, and
$R_{\rm ip}$,
the inflexion radius where $\frac{d^2 I}{dR^2}=0$ are shown by
vertical dashed lines
(left) and circles (right) of blue and red colours, respectively.
Note that the profiles are shown with logarithmic steps to illustrate
the inner and
outer profile shapes, simultaneously.}
\label{fig:profMD8}
\end{center}
\end{figure*}

\begin{table}
\begin{center}
\caption{Fitting radius and background value for all M82 disk SSCs}\label{tab:rfit}
\begin{tabular}{lrrrrr}
\hline
ID$^\dag$ & R.A. & DEC & \Rip & $R_{\rm 3\sigma}$& $\mu_{\rm bg}\pm \delta\mu$\\
 & (deg) & (deg) & (pix)  & (pix) &  ($\rm mag/arcsec^2$) \\
 (1) & (2) & (3) & (4) & (5) & (6)\\
\hline
   D1    &         148.94615        &	69.67842	&	19 &            23 &          18.42$\pm$0.22\\
   D2    &         149.00384        &	69.68513	&	14 &            24 &          18.20$\pm$0.30\\
   D3    &         149.01420        &	69.68672	&	18 &            50 &          18.89$\pm$0.44\\
   D4    &         148.94655        &	69.67856	&	9   &            26 &          18.48$\pm$0.22\\
   D5    &         148.94480        &	69.67737	&	11 &            16 &          19.11$\pm$0.32\\
   D6    &         149.00952        &	69.68584	&	18 &            34 &          18.74$\pm$0.32\\
   D7    &         149.00372        &	69.68566	&	11 &            14 &          17.84$\pm$0.17\\
   D8    &         149.05942        &	69.69909	&	25 &            26 &          20.34$\pm$0.12\\
   D9    &         148.98191        &	69.68497	&	14 &            34 &          19.02$\pm$0.21\\
  D10    &        148.98767        &	69.68537	&	9 &            28 &          18.88$\pm$0.16\\
 \hline
\end{tabular}
\end{center}
\hfill\parbox[t]{\columnwidth}{$^\dag$IDs from \citet{Mayyacat}.  The 'D' preceeding the numbers stands for 'disk' sample}
\end{table}

\begin{figure*}
\includegraphics[width=\textwidth]{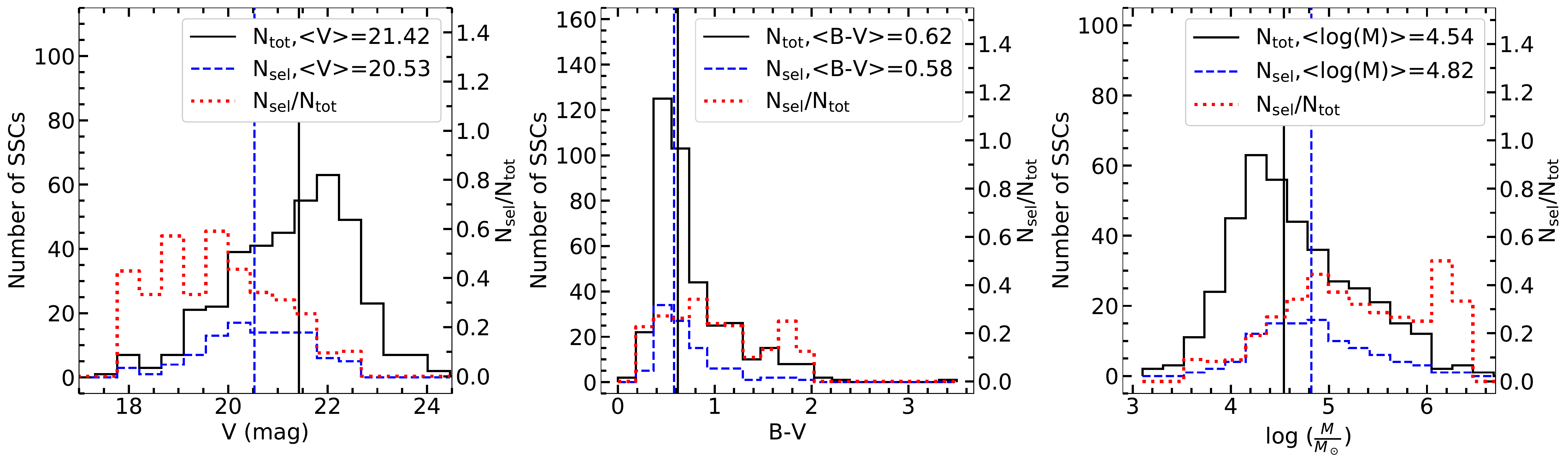}
\caption{Comparison of distributions of V Magnitude (left), $B-V$ (middle) and Cluster mass (right) of all M82 disk SSCs (black solid line) with the selected sample (99 clusters) (blue dashed line).  Median values of the distributions are shown by the vertical lines and written in the top-right corner.  The fraction of objects ($N_{\rm sel}/N_{\rm tot}$) in each bin is shown by a red dotted line.}
\label{fig:mag_color_av_sub}
\end{figure*}

Profile fitting packages such as GALFIT and ISHAPE analyse 2-dimensional (2-D) images to
obtain structural parameters of star clusters. On the other hand, azimuthally
averaged 1-dimensional (1-D) profiles have been used traditionally to obtain structural
parameters of well-resolved clusters in the Milky Way and local group galaxies 
\citep[e.g.][]{Elson, McLaughlin2005}.
The profile functions used in both these techniques are 1-D profiles applicable to 
spherically symmetric 3-D models. Moreover, derived structural parameters such as core radius
and half-light radius refer to the radially symmetric configurations. 
We hence adopt the 1-D technique in this study. The {\it ellipse}  task 
\citep{Ellipse_iraf1987} in IRAF/STSDAS package is the standard tool to obtain the
azimuthally averaged 1-D profiles of extended objects. Though the task is developed
for analysing the surface brightness profiles of external galaxies, the task does an
excellent job in obtaining 1-D SBPs of clusters on the HST images.
We fixed the centers of the ellipses at the centroids of the SSCs, and obtained the
SBPs at successive HST/ACS pixels, with a width of 1 pixel (0.88~pc at the distance of M82).
The task also calculates the azimuthal dispersions in the intensity at each radial bin,
which is a measure of errors in the SBPs.
We left the ellipticity ($\epsilon=1-b/a$, where $a$ and $b$ are semi-major and 
semi-minor axes, respectively) and position angle as free parameters in
a first run of the task.  
The distribution of ellipticities peaks at 0.19 with only 25\% of the 
SSCs having higher ellipticities.  This is illustrated in Fig. \ref{fig:profellip} of the Appendix.    Thus,
the majority of the SSCs are nearly circular.  With $\epsilon=0.3$, the well-known SSC M82-F (D1) is one of the most elongated clusters.
The SBP obtained for this cluster using circular and elliptical ($\epsilon=0.30$)
apertures are similar (see Fig. \ref{fig:profellip2} in the Appendix).
\citet{SmithGallagher2001} also found that the half-light radius obtained from profiles using 
circular and elliptical apertures for M82-F are similar, which confirms
that the derived structural parameters are not very sensitive 
to small differences in ellipticities.
We hence fixed the ellipticity at the minimum value permitted by the task, 
which is $\epsilon=0.05$.
     
The surface brightness profiles obtained by the {\it ellipse} task contain 
background contribution which should be subtracted in order to obtain pure 
cluster profiles.  This background in the case of M82 clusters mainly comes 
from its disk, which varies appreciably from one cluster to another.  
This makes the measurement of background for each cluster mandatory.
We analysed the four corners of the cut-outs for finding an appropriate local 
background value for each SSC.  Median and standard deviation values were 
obtained in boxes of $10\times10$ pixel size at the four corners of the cut-out
images, the minimum of these four values being chosen as the background 
$I_{\rm bg}$ and the noise $\sigma_{\rm bg}$ respectively.  
For each background-subtracted profile, we determined a limiting radius, defined 
as the outer-most point of the profile at which the cluster surface brightness 
is equal to 3 times $\sigma_{\rm bg}$. We refer to this radius as \Rbg. 

Cluster SBPs are expected to monotonically decrease up to \Rbg. However, we found
that the majority of the profiles have an inflexion point at $R<$\Rbg.
Visual examination of the images suggested that this is due to stars or clusters
in the neighbourhood of the object of analysis. When
possible, we masked the contaminating sources in each cut-out image before obtaining
SBPs.
In a few cases, the masks were successful in producing SBPs free from 
contamination from neighbours. However, in the majority of the cases, SBPs are 
affected due to some residual contribution from the neighbours. This is because 
in most of the cases the contaminating source is another SSC, which occupies a non-negligible
number of pixels of the cut-out image.   We took into account this effect by 
defining a fitting radius for each SSC, within which the profile is free from 
contamination from a neighbour. We obtained this radius by determining the innermost inflexion point 
\Rip, such that at \Rip, $\frac{d^2 I}{dR^2}=0$, for each background-subtracted 
profile.  
In general, the fitting radius, \Rfit, is the minimum of \Rip\ and \Rbg.  
However in all cases \Rip\ $<$ \Rbg, and hence, \Rfit $=$\Rip.

The profile analysis was carried out in the BVI HST/ACS bands.  In each of 
these bands, the above procedure is repeated.  Hence, we have a set of \Ibg, 
\sigmabg, \Rip, \Rbg,  and \Rfit\ for each band.

In Figure \ref{fig:profMD8}, we illustrate the procedure adopted for obtaining the fitting radius for 
two clusters, one unaffected by a contaminating object (D8) and the 
other with a bright nearby contaminating source (D4). In the former case, \Rip\ is almost equal to \Rbg, whereas in the latter case,  \Rip\ is less than half of \Rbg. The bump immediately beyond \Rip\ is due to the contaminating object, which can be seen in the RGB image.  In Table \ref{tab:rfit}, we give the values of \Rfit, \Rbg\ and the background surface brightness ($\mu$) and its error  ($\delta\mu$), as well as the R.A. and DEC for all SSCs.
The error is calculated as $\delta\mu=1.086\left(\frac{\sigma_{\rm bg}}{I_{\rm bg}}\right)$.  \Rfit\ in all cases is the inflexion radius \Rip.  

In the following subsection, we will discuss the global properties of the selected 99 SSCs with respect to the total SSCs disk sample.  

\subsection{Selection and characteristics of the disk subsample}\label{Sec:gen_char}

In order to obtain reliable structural parameters from the HST images, star clusters have to satisfy certain criteria.  
The most important of these criteria is that they have enough number of pixels for profile analysis.  
In the absence of a contaminating source, the number of pixels for analysis depends on the intrinsic size of the cluster.
Another criterion for selection of clusters is that the extracted profile is well-fit by one of our models, quantified by $\chi^2$ statistics, and described in the next section. 
We carried out an analysis of synthetic clusters in order to define the \Rfit\ necessary to reliably recover the input parameters, which is also described in the next section.  
Based on this analysis, we considered a cluster to be good for analysis (1) if the \Rfit~$\ge 8$\,pixels 
in at least two bands and (2) the $\chi^2$ of the best fit is less than 3$\times\nu$, where $\nu$ is the number of degrees of freedom \citep{Avni1976,Wall_statistics}  (criterion 1 and 2, respectively). In the majority of the cases, B and V bands have similar behaviour with the I band displaying a different behaviour, with $\sim$16\% having less than 8 pixels in I.  
The imposition of this criterion reduced our sample size to a subsample of 99 SSCs.     

In order to determine how representative is the subsample with respect to the 
total sample, we compare the distributions of three of the most important 
characteristics for the two samples in Fig \ref{fig:mag_color_av_sub}.
The chosen characteristics are: $V$ magnitude, $B-V$ colour and the photometric mass.  
Given that most of the disk SSCs are formed in a disk-wide burst around 300~Myr ago, little dispersion is expected in the intrinsic colours and mass-to-light ratios of the SSCs.
However, M82 SSCs suffer from considerable extinction, which gives rise to large dispersion in their colours.  
In this work, we assume that the entire dispersion in the colour histogram is caused by reddening.  
Hence, extinction-corrected magnitude is directly a measure of the mass for these SSCs.
Our subsample contains between 20--40\% of the total sample for B-V=0.2--2.0~mag, V=18--22~mag, and $\log{M/M_\odot}$=4.5--6.5~dex.
In summary, our subsample represents the bright (V>18.0~mag), massive (M>$3\times 10^4 M_\odot$) end of the total sample of SSCs, covering uniformly the entire range of extinction values.
This subsample is complete above mass of $M>3\times 10^4 M_\odot$, which is very close to the turn-over in the mass function for the entire cluster sample \citep{Mayyacat}.  
Thus, our subsample is representative of the massive end of the luminosity function.  

\section{Determination of structural parameters}\label{Sec:method_fit}

Structural parameters were obtained by fitting the observed SBPs with 
PSF-convolved theoretical profiles.
The fitted model profiles are Moffat-EFF \citep{Elson}, King \citep{King_dyn}, 
and Wilson \citep{Wilson_dyn}. We followed the procedure described in detail 
in \cite{McLaughlin2005}, which we summarise briefly in this section.

\subsection{Dynamical models of star clusters}

King and Wilson models are based on a ``lowered'' Maxwellian  kinetic energy distribution function of stars.   
These two model structures differ only in the outer halo regions, which is due to an extra term in the Wilson formulation and 
are defined in terms of the distribution function of a relative energy $\mathcal{E}=-E+\Phi_0$, where $E$ is the total energy for a star moving with an isotropic velocity $v$ under a potential $\Phi$.   The term $\Phi_0$ is a constant such that the relative energy is positive everywhere in the cluster.  Under this formulation,  the relative potential $\Psi=-\Phi+\Phi_0$ and $\mathcal{E}=\Psi-\frac{1}{2}v^2$.  The models described above are given by:

\begin{equation}
{\rm King:}  f(\mathcal{E}) \propto \left\{ 
\begin{array}{cc}
e^{(\mathcal{E}/\sigma_0^2)} -1, & \mathcal{E}>0,\\
0, & \mathcal{E} \leqslant 0,   
\end{array}
\right. ,
\label{distfuncking}
\end{equation}

\begin{equation}
{\rm Wilson:}  f(\mathcal{E}) \propto \left\{ 
\begin{array}{cc}
e^{(\mathcal{E}/\sigma_0^2)} -1-\frac{\mathcal{E}}{\sigma_0^2}, & \mathcal{E}>0,\\
0, & \mathcal{E} \leqslant 0,   
\end{array}
\right. ,
\label{distfuncwilson}
\end{equation}
where $\sigma_0$ is a scale parameter which 
measures the core dispersion velocity defined as: 

\begin{equation}
\sigma_0^2 \equiv \frac{4\pi G \rho_0 r_0^2}{9},
\label{relkingsigma}
\end{equation}
where $\rho_0$ is the central stellar density and $r_0$ the scale radius,
commonly referred to as King radius. 
These models are parametrised by a dimensionless potential $W=\Psi/\sigma_0^2$,
which is defined at all radii inside the tidal radius $r_{\rm t}$, and has
the boundary values of $W(r=0)\equiv W_0$ and $W(r=r_t)=0$. $W_0$ is a measure
of the central potential, being directly related to the often-used
concentration parameter $c=\log(\frac{r_t}{r_0})$. In this work, we varied the $W_0$
values between 2 and 15, which corresponds to $c=$0.5 and 3.3 for King models 
and $c=$0.7 and 4.1 for Wilson models.

\begin{figure*}
\begin{center}
\subfloat{\includegraphics[width=0.31\textwidth]{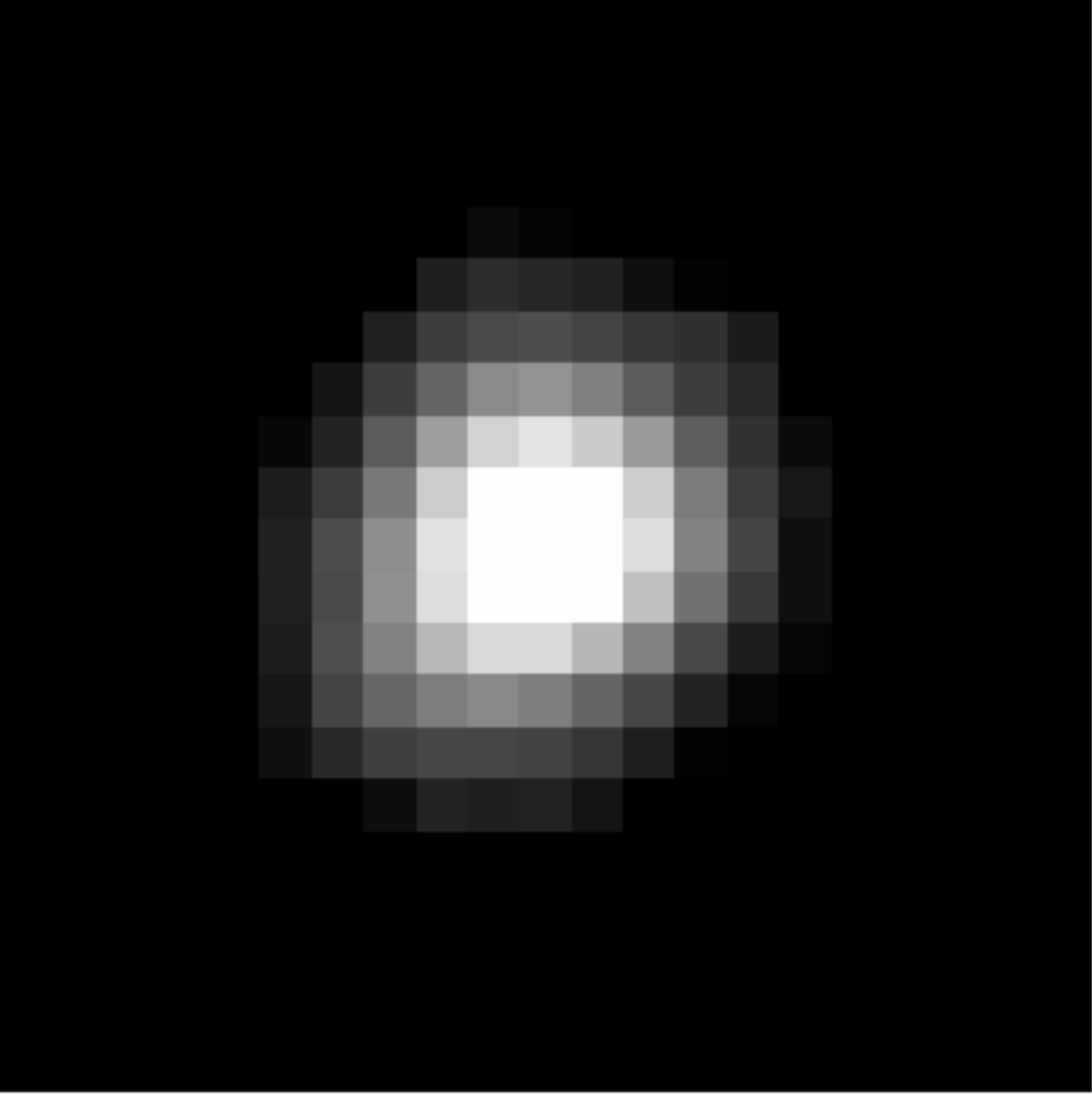}}
\subfloat{\includegraphics[width=0.31\textwidth]{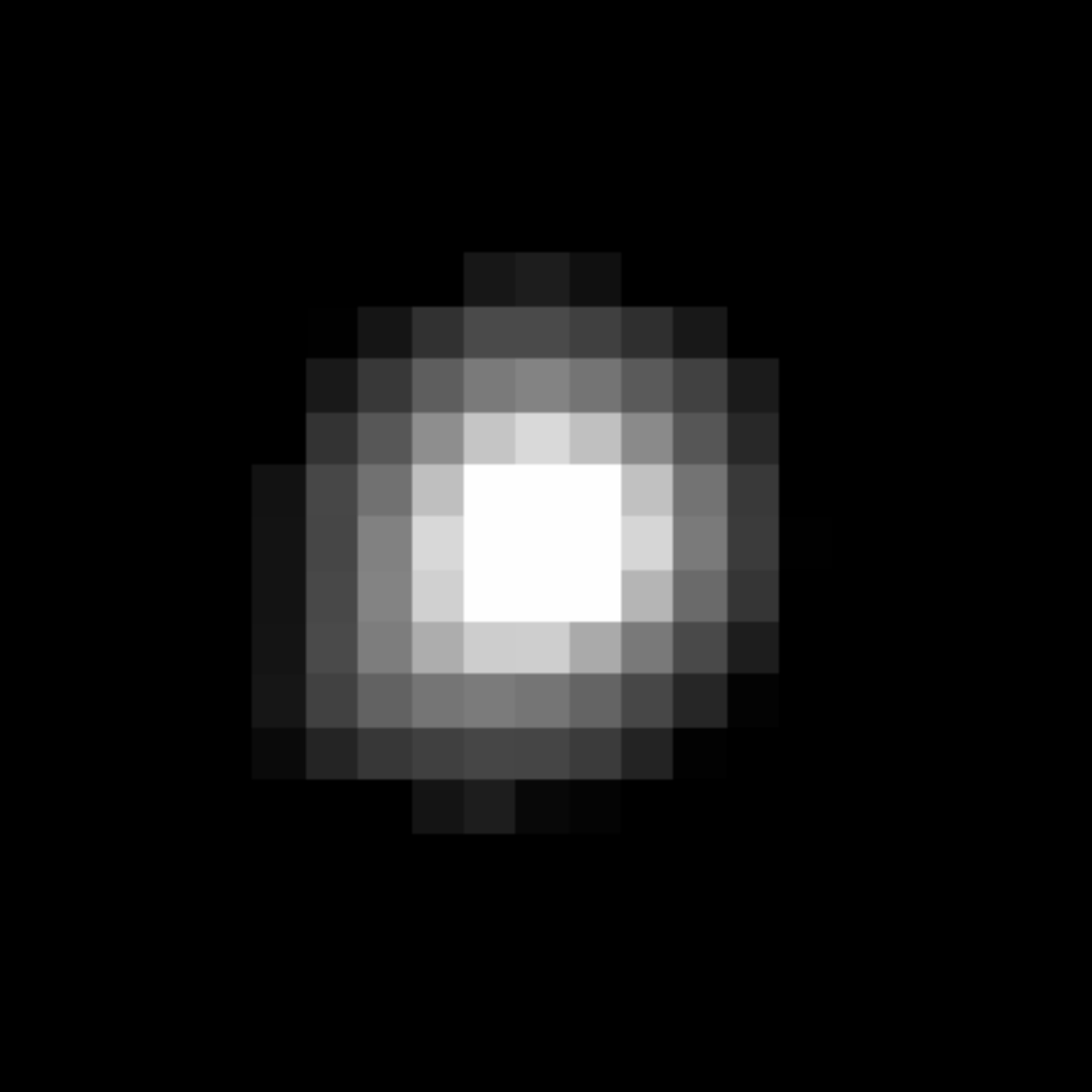}}
\subfloat{\includegraphics[width=0.31\textwidth]{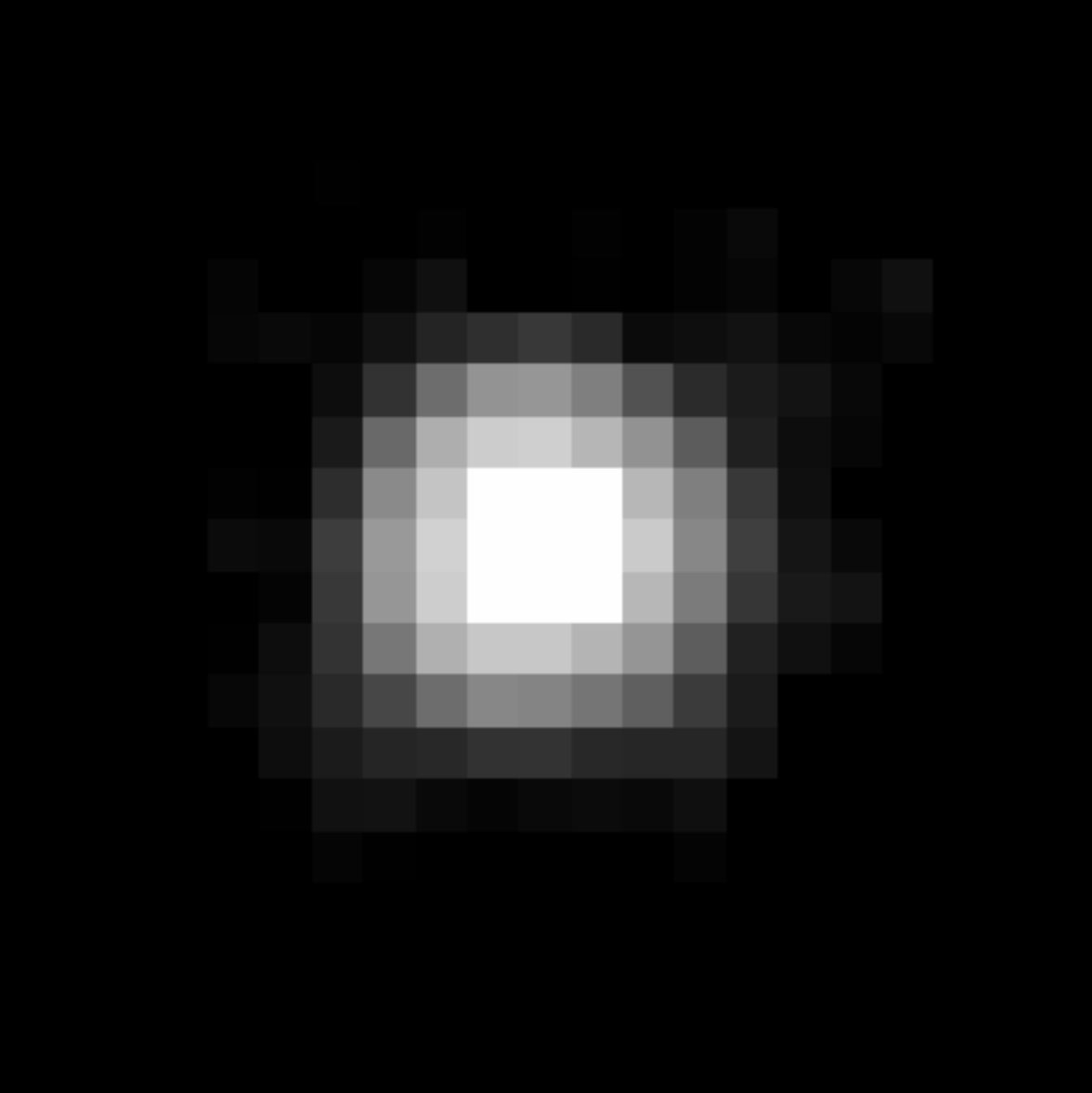}}

\caption{PSF images in B (left), V (middle) and I (right) bands calculated using PSFex \citep{Bertinpsfex}.}
\label{fig:psf_indiv}
\end{center}
\end{figure*}

The solution of these equations is expressed as a function of $W(r)$, which is 
directly related to the 3-D density function, $\rho(r)$ through the Poisson equation (as well as a normalised velocity dispersion profile, in terms of the central velocity dispersion, solving Jeans Equation). The observable quantity $I(R)$ is obtained from $\rho(r)$ by projecting it into the plane
of the sky along the $R$ axis following the standard formulation \citep[e.g. Eq. 2.138a in][]{Galdynbook} and dividing by the mass-to-light ratio $\Gamma$:

\begin{equation}
I(R) = \frac{\Sigma(R)}{\Gamma} = \frac{2}{\Gamma} \int_R ^{R_t} \frac{\rho(r)}{(r^2-R^2)^\frac{1}{2}} rdr,
\label{rho_surf_king_eq}
\end{equation}
where the integration limits are defined as $R=r/r_0$ and $R_t=r_t/r_0$, $r_0$
being obtained by fitting model profiles to observed SBPs.
The cluster extends up to the tidal radius $r_{\rm t}$, where by definition $E=0$. 

The Moffat-EFF profiles were proposed by \cite{Elson} as a convenient 
modification of the empirical King profile \citep{King_emp} to fit the SBPs of 
LMC clusters. The functional form of the profile is:

\begin{equation}
I(R)=\frac{(\gamma-2) L_{\rm tot}}{2 \pi r_{\rm d}^2} \bigg{[}1+\bigg{(}\frac{R}{r_{\rm d}}\bigg{)}^2\bigg{]}^{-\gamma/2}, 
\label{eq:moffat_proj}
\end{equation}
where $R$ is the semi-major axis of the observed profile, $r_{\rm d}$ is the 
characteristic radius which is related to the core radius, $R_{\rm c}$ by:

\begin{equation}
r_{\rm d}=\frac{R_{\rm c}}{(2^{2/\gamma}-1)^{1/2}}
\label{rc_rd}
\end{equation}
Once $\gamma$ and $r_{\rm d}$ are determined from the fitting, the 3-D luminosity density
profile can be calculated using the expression:

\begin{equation}
j(r)= j_{\rm 0} \bigg{(}1+\frac{r^2}{r_{\rm d}^2}\bigg{)}^{-(\gamma+1)/2},
\label{moffat_eq}
\end{equation}
The mass density is obtained using $\rho(r)=\Gamma j(r)$.  On the other hand, a velocity profile is found solving Poisson and Spherical Jeans equations, giving rise to a normalised velocity dispersion profile in terms of the central velocity dispersion $\sigma_0$.  The surface mass-density $\Sigma$ is found by projecting the volume mass density $\rho(r)$ into the plane of the sky, following Eq. \ref{rho_surf_king_eq}.  The surface density profile also allows us to calculate numerically $R_{\rm h}$, the radius containing half the total light.  

\begin{figure}
\begin{center}
\includegraphics[width=\columnwidth]{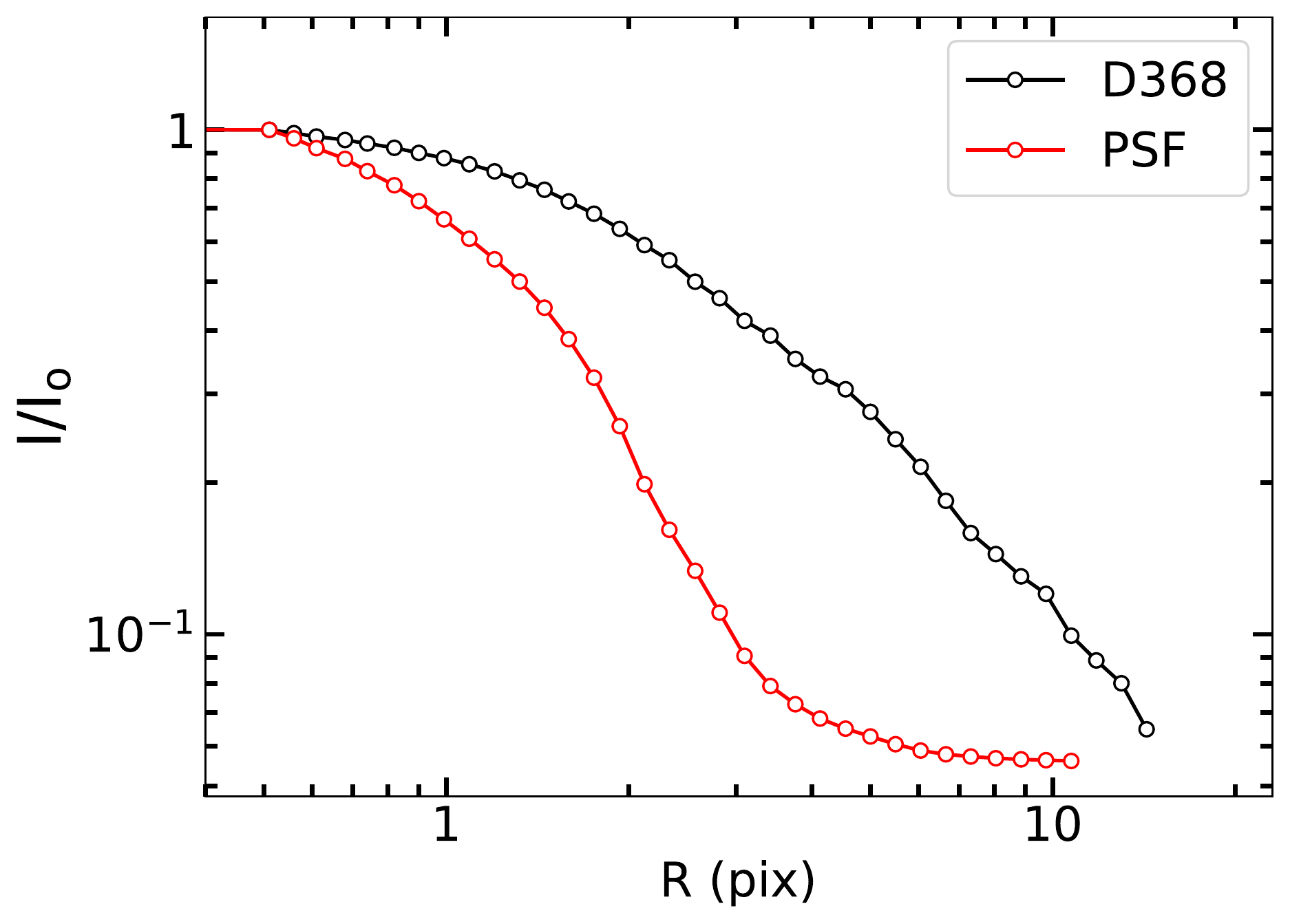}
\caption{PSF profile (red) compared to that for one of the most compact clusters of the sample (D368: black), both normalised to their peak values in the $V$-band.  These radial profiles are obtained using the IRAF task \emph{ellipse}, at logarithmic steps increasing successively by 10\%.  The cluster profiles are clearly broader than the PSF profiles.}
\label{fig:psf_prof}
\end{center}
\end{figure}

\subsection{PSFs}\label{Sec:psf_des}

Intrinsic cluster profiles are broadened due to the PSF of the instrument, and
hence in order to determine the structural parameters, especially the core
parameter, it is essential to convolve the model profiles with the instrumental PSF profiles 
before comparing with observed SBPs.
We used the PSFex \citep{Bertinpsfex} tool to obtain a PSF in each of the three bands.  Before using this tool, we selected a list of suitable stars in each of these bands using SExtractor \citep{Bertinsextractor}.  A star is considered to be suitable for PSF construction if it is isolated, and bright, but not saturated.  We used various SExtractor output parameters to select these PSF stars. More than 1000 stars 
were used in each of the bands for this purpose (1234 in B, 1401 in V and 1328 in I).  The resulting PSFs are shown in Fig. \ref{fig:psf_indiv}.      
 
In order to illustrate that our selected SSCs are easily distinguishable from stars, we compare in Fig \ref{fig:psf_prof} the profile for one of the smallest SSCs with that of the PSF.  Each displayed profile is an azimuthally averaged profile obtained as described in \S \ref{Sec:sbps}.  In this illustration, both profiles are sampled in logarithmic steps successively increasing by 10\%.  Profiles generated for fitting purposes have linear steps of 1 pixel size for clusters and 0.5 pixel size for the PSF.  

\subsection{$\chi^2$ method and errors on derived parameters}\label{Sec:chi_an_errs}

In order to extract structural parameters from SBPs, we used the $\chi^2$-minimisation technique. We define $\chi^2$ as: 

\begin{equation}
\chi^2=\sum_{i=1}^{N{\rm pts}} \frac{(I_{{\rm obs}_i}-\tilde{I}_{{\rm model}_i})^2}{\sigma_i^2},
\label{eq:chi_sq}
\end{equation} 
where  $I_{{\rm obs}_i}$ and $\tilde{I}_{{\rm model}_i}$ are the i$^{\rm th}$ point in the observed SBP and PSF-convolved model profile, respectively.  The summation is over     
$N$pts, within the fitting radius, \Rfit.   The $\sigma_i$ term is the $I_{\rm err}$ of the azimuthally averaged i$^{\rm th}$ isophote, as calculated by the {\it ellipse} task.  

The convolution of the model with the PSF was performed with the Fortran routine CONVLV from {\sc Numerical Recipes} \citep{Numrec} which performs a FFT.   For this purpose, we sampled both the PSF and the model at linear steps of 0.5 pixels which is two times the sampling of the objects.

The $\chi^2$ fitting technique is implemented using a Fortran program developed for this purpose.  The best-fitting parameters were obtained using a two-step procedure:  in the first step, we used a coarse grid in the parameter space to obtain a preliminary minimum $\chi^2$.  In the second step, we used ten-times better steps  and searched for minimum  $\chi^2\equiv \chi^2_{min}$ around the preliminary parameters set, to cover a range of  four-times the coarse step.  
In Tab. \ref{tab:fit_details}, we give the range and the coarse step size for the parameters. The fitting procedure starts with coarse grids.  Once a local minimum is found, the fine grid is used.  We repeated the second step around the next two local minima of the first step. In all cases, the best-fit parameter set using fine steps is around the values corresponding to the minimum $\chi^2$.  This two-step procedure ensures that the recovered parameters have a numerical precision better than the coarse step. This also resulted in a parameter set of nearly one thousand models that satisfy the condition $\chi^2-\chi_{\rm min}^2<1$. 

\begin{table}
\begin{center}
\caption{Range and step of parameters values for the three fitted models}
\label{tab:fit_details}
\begin{tabular}{crr}
\hline
Model & $r_{\rm d}$ or $r_{\rm 0}$ & $\gamma$ or $W_{\rm 0}$\\ 
(1) & (2) & (3)\\ 
\hline
Moffat-EFF & 0.05--10;0.04 & 0.1--10;0.04 \\
King & 0.05--10;0.10 & 2.0--15;0.10\\
Wilson & 0.05--10;0.10 & 2.0--15;0.10\\
\hline
\end{tabular}
\end{center}
\hfill\parbox[t]{\columnwidth}{Col (1): Model.  Col (2): Moffat-EFF $r_{\rm d}$ or Dynamical (King or Wilson) $r_{\rm 0}$ ranges and step size in pixels.  Col (3): Moffat-EFF $\gamma$ parameter or Wilson or King $W_{\rm 0}$ parameter range and step size.} 
 \end{table}

\begin{figure}
\begin{center}
\includegraphics[width=\columnwidth]{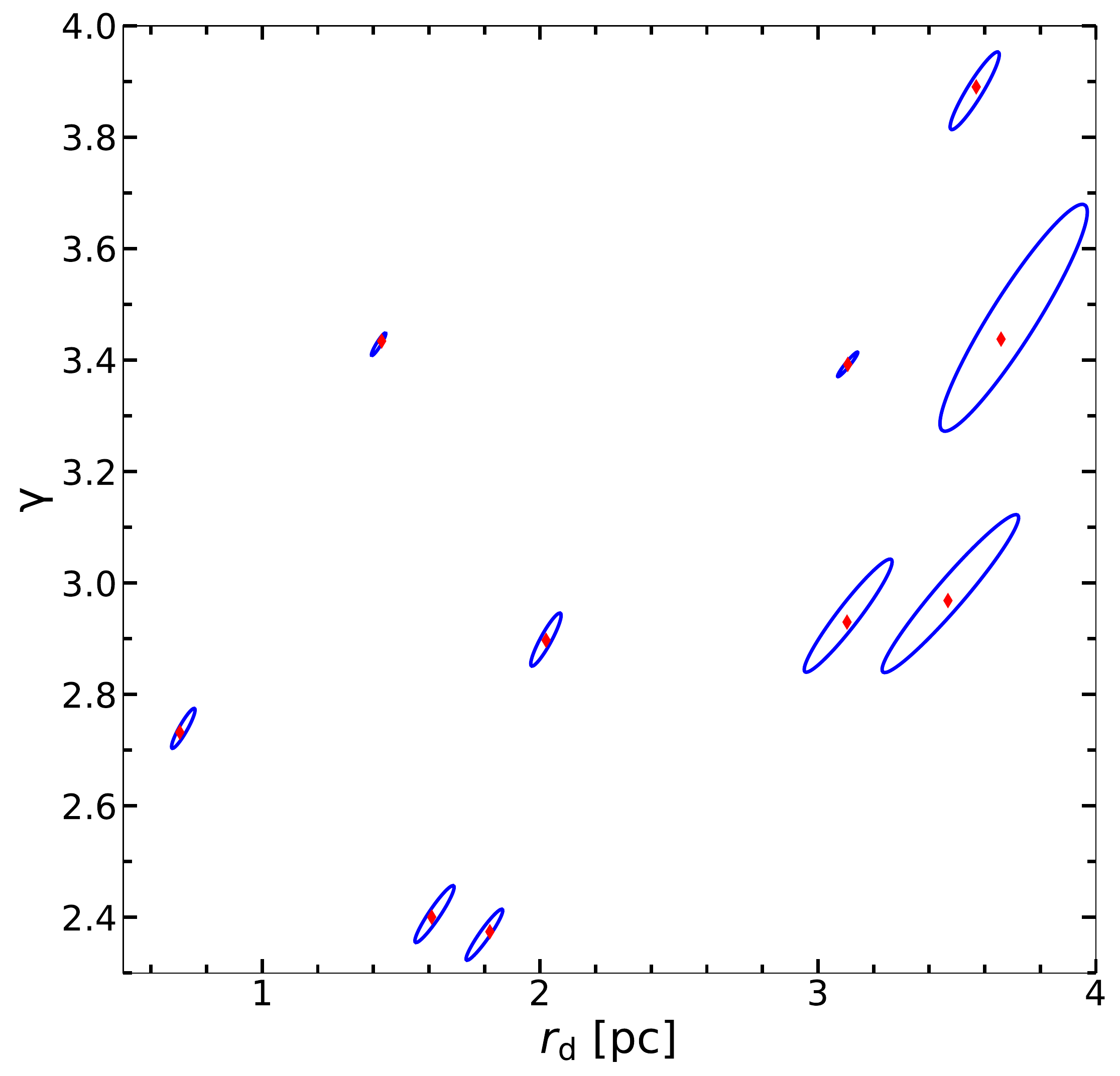}
\caption{Error ellipses (1-$\sigma$ confidence intervals) of the parameters $r_{\rm d}$ and $\rm \gamma$ of models satisfying the criterion $\chi^2-\chi_{\rm min}^2<1$ for the brightest 10 observed clusters of the sample.  The errors on both axes correspond to the projections of the ellipse along x-axis and y-axis.   The best-fitted  parameters are shown with red diamonds.}
\label{fig:conf_el}
\end{center}
\end{figure}

\begin{figure}
\begin{center}
\includegraphics[width= \columnwidth]{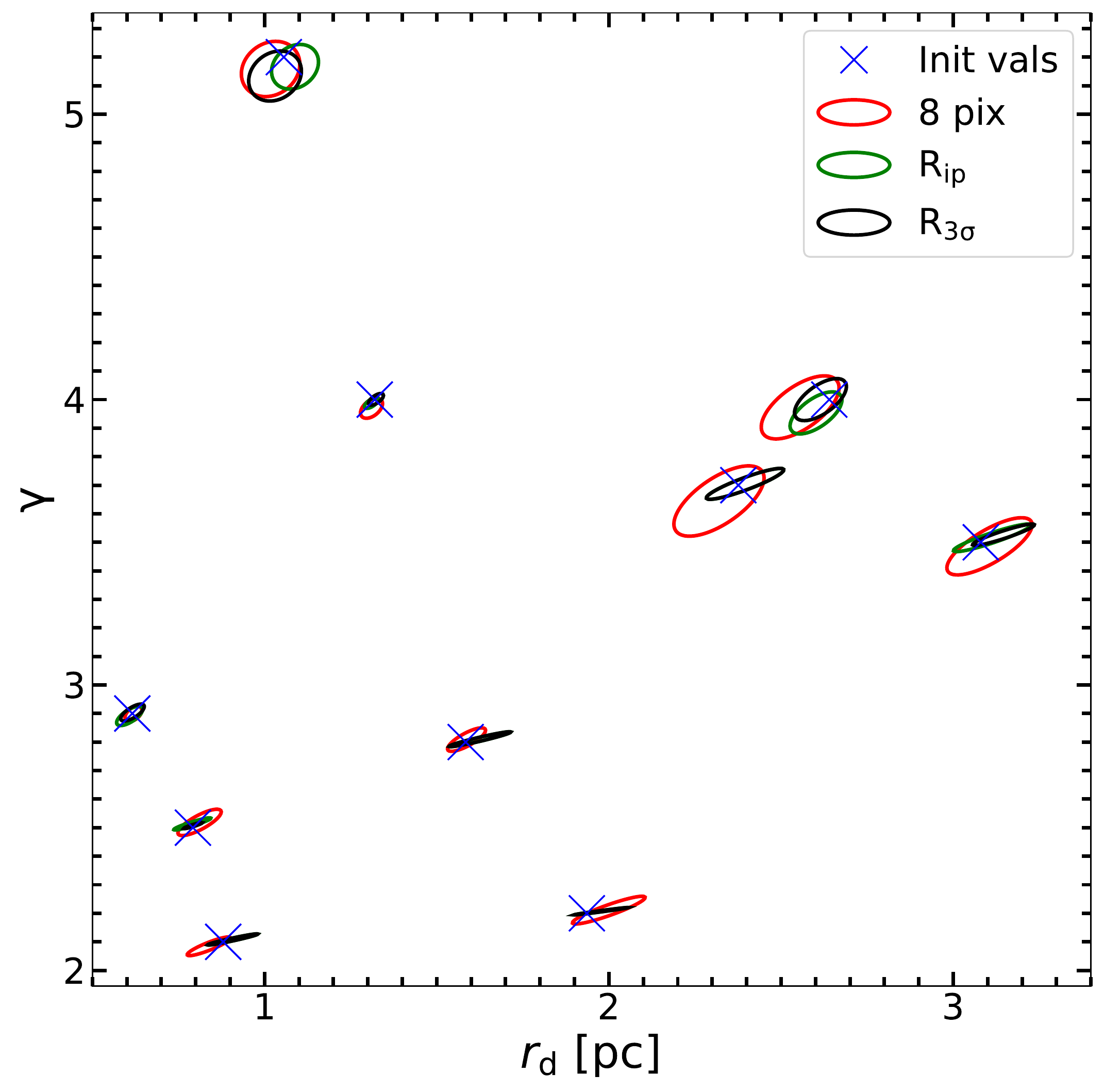}
\caption{Error ellipses (1-$\sigma$ confidence intervals) of the parameters $r_{\rm d}$ and $\rm \gamma$ of models satisfying the criterion $\chi^2-\chi_{\rm min}^2<1$ for 10 synthetic clusters  that mimic the properties of the clusters of the sample.  For each synthetic cluster, parameter values for 3 \Rfit\ values are shown: \Rfit=8~pixels (red), \Rip\ (green) and $R_{\rm 3 \sigma}$ (black).  The initial values are shown by crosses, which are inside the error ellipses even with the \Rfit=8~pixels.}
\label{fig:comp_rip}
\end{center}
\end{figure}

\begin{figure*}
\begin{center}
\includegraphics[width= \textwidth]{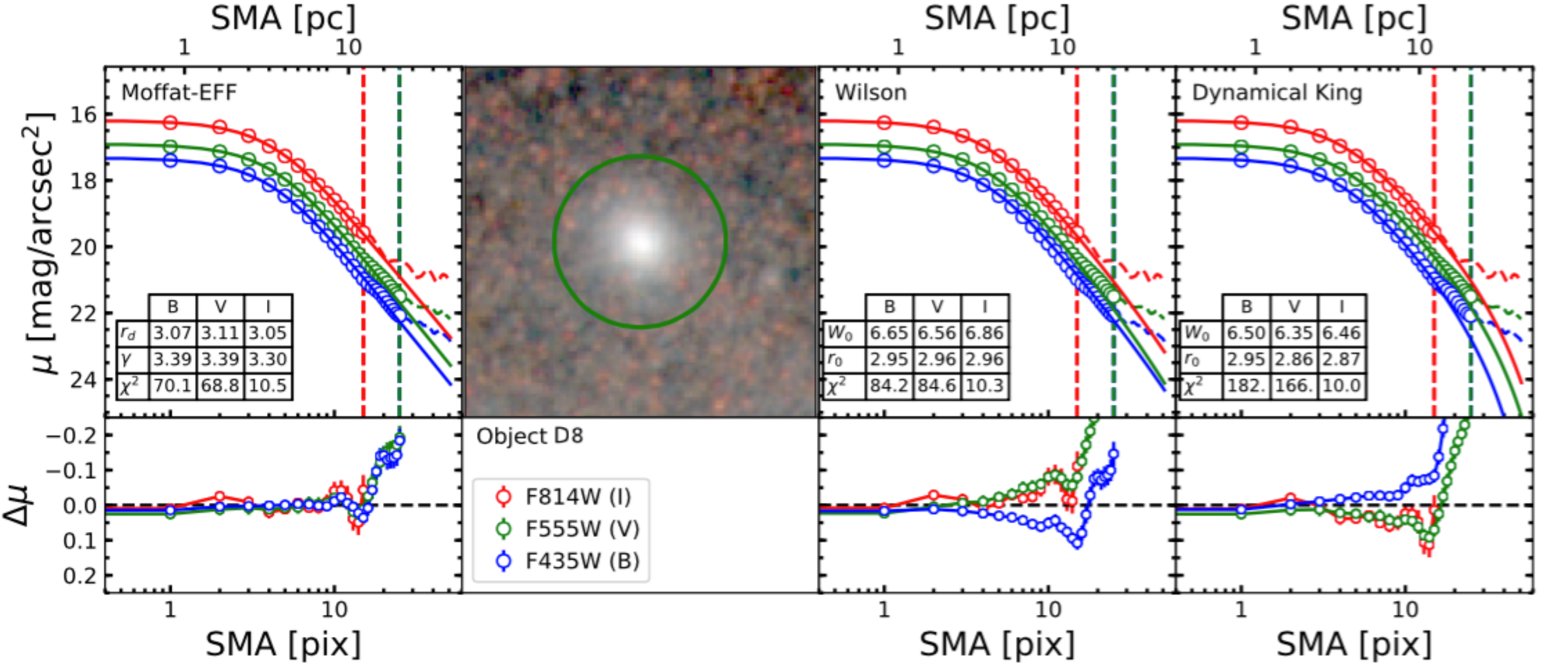}
\caption{Illustration of the dynamical model fitting of the observed profiles with Moffat-EFF (left-most), Wilson (third from left) and Dynamical King (right-most). Profiles for SSC D8 in the filters B (blue), V (green) and I (red) are shown, and the fitting radius in the corresponding filter is indicated with dashed vertical lines following the same colour code.  In this particular case, the \Rfit\ corresponding to filters B and V coincide.  In the bottom panels, the corresponding residuals ($\Delta \mu= \mu_{\rm obs} -\mu_{\rm model}$) are shown. An RGB cut-out image of the cluster formed using I(R), V(G) and B(B) band images is shown, with a circle indicating the \Rfit\ of the V-band.  Each fitted point is shown by circles for radius $\leq$ \Rfit, and by dashed lines beyond that.  The error bars are shown both in the fittings and the residuals, but are smaller than the symbol size for the great majority of points. The best-fit parameter values and the  $\chi^2_{\rm min}$ for each fit are shown in the embedded tables.  The electronic version contains plots such as this for all the 99 SSCs.}
\label{fig:fittings_all}
\end{center}
\end{figure*}

The errors on the best-fit parameters for all 99 SSCs were obtained based on the $\chi^2$ statistics.  We considered parameter values of all our models for which $\chi^2-\chi_{\rm min}^2<1$, as a set of acceptable values within 1-$\sigma$ confidence limit.  We show this set of parameters in $\gamma$ vs $r_{\rm d}$ plane as error ellipses  for 10 of our objects for the Moffat-EFF profile fits in Fig. \ref{fig:conf_el}.  The error on $r_{\rm d}$ and $\gamma$ correspond to the projections of the ellipse along the x-axis and y-axis, respectively.  We would like to note that the best-fit parameters are not necessarily at the centre of the ellipse, implying the errors on the positive and negative sides are not always the same.  A clear tendency is seen for errors being larger for larger values of the parameters.   These 1-$\sigma$ confidence limits are reported as the errors on $r_{\rm d}$ and $\gamma$ for all the clusters in Tab. \ref{tab:basic_fit_pars}. A similar analysis was carried out to obtain the errors on the parameters of King and Wilson models, which are also reported in the table.        

\subsection{Fits on simulated clusters and minimum \Rfit}

As mentioned in \S\ref{Sec:sbps}, the choice of \Rfit\ is crucial in determining reliable structural parameters.  With the aim of determining the minimum number of pixels required for this, we carried out a profile fitting procedure on simulated clusters.  The mock sample of clusters constituted 10 clusters, all following Moffat-EFF profiles and covering the extreme ranges of the parameter space. An rms noise is added to the simulated images, which are also convolved with the PSFs described in \S\ref{Sec:psf_des} in order to simulate the observational effects.   The structural parameters of the mock sample were recovered following the same procedure as that for the sample clusters.  For each cluster, we obtained structural parameters for several values of \Rfit, starting at 6~pixels, and all the way up to \Rfit =$R_{\rm 3\sigma}$.  

For each of these \Rfit\ values, we obtained the best-fit parameters as well as their error ellipses.
We found that for \Rfit$<8$~pixels, $\chi^2_{\rm min}$ values in general were greater than $3\times\nu$ and hence do not satisfy our selection criteria.    
In Fig. \ref{fig:comp_rip}, we show the results of the simulations in $\gamma$ vs $r_{\rm d}$ plane for three values of \Rfit: 1. \Rfit =$R_{\rm 3\sigma}$ ,  2. \Rfit =\Rip, and 3. \Rfit = 8~pixels.  As expected, \Rfit =$R_{\rm 3\sigma}$ has the least error, with the maximum error being for the \Rfit=8~pixels.  Even in the latter case, the recovered values are in good agreement with respect to the initial values.  Thus, we conclude from these simulations that the parameters values recovered with \Rfit = 8 pixels for our observed sample are reliable.

\begin{table*}
\caption{Best-fit parameters in filter V for Moffat-EFF, King and Wilson models.}
\label{tab:basic_fit_pars}

\begin{tabular}{cccrrrrrr}
\hline
ID & Npts & Model & $\chi^2_{\rm min}$ & $W_{\rm 0}$ or $\gamma$ & $r_ {\rm 0}$ or $r_{\rm d}$ & $\mu_{\rm 0}$ & $\log{I_{\rm 0}}$\\
 & & & & & (pc) & (mag/arcsec$^2$) & ($L_\odot$/pc$^2$)\\
(1) & (2) & (3) & (4) & (5) & (6) & (7) & (8)\\ 
\hline
   D1  & 19 & 		M & 	 12.69 & 	 2.73$_{-0.03}^{+0.04}$ & 	 0.70$_{-0.03}^{+0.06}$ & 	 13.55$_{-0.06}^{+0.11}$ & 	  5.16$_{-0.04}^{+0.07}$ \\
 & & 	K & 	 22.92 & 	 8.70$_{-0.05}^{+0.04}$ & 	 0.92$_{-0.19}^{+0.10}$ & 	 13.55$_{-0.28}^{+0.15}$ & 	  5.16$_{-0.20}^{+0.11}$ \\
 & & 	W & 	 10.04 & 	 8.76$_{-0.01}^{+0.08}$ & 	 0.84$_{-0.10}^{+0.19}$ & 	 13.55$_{-0.14}^{+0.29}$ & 	  5.16$_{-0.10}^{+0.21}$ \\
  D4  & 9 & 		M & 	 11.57 & 	 2.97$_{-0.13}^{+0.15}$ & 	 3.47$_{-0.27}^{+0.29}$ & 	 16.50$_{-0.42}^{+0.46}$ & 	  3.98$_{-0.30}^{+0.33}$ \\
 & & 	K & 	 11.65 & 	 7.86$_{-1.36}^{+7.14}$ & 	 3.56$_{-0.10}^{+0.40}$ & 	 16.50$_{-1.93}^{+10.1}$ & 	  3.98$_{-1.36}^{+7.15}$ \\
 & & 	W & 	 11.72 & 	 8.95$_{-1.55}^{+6.05}$ & 	 3.56$_{-0.10}^{+0.40}$ & 	 16.50$_{-2.20}^{+8.57}$ & 	  3.98$_{-1.55}^{+6.06}$ \\
  D7  & 11 & 		M & 	  1.46 & 	 2.37$_{-0.05}^{+0.04}$ & 	 1.82$_{-0.10}^{+0.05}$ & 	 16.85$_{-0.15}^{+0.09}$ & 	  3.84$_{-0.11}^{+0.07}$ \\
 & & 	K & 	 18.86 & 	14.96$_{-2.56}^{+0.04}$ & 	 2.52$_{-0.10}^{+0.19}$ & 	 16.85$_{-3.62}^{+0.28}$ & 	  3.84$_{-2.56}^{+0.20}$ \\
 & & 	W & 	 19.42 & 	14.96$_{-1.46}^{+0.04}$ & 	 2.52$_{-0.10}^{+0.19}$ & 	 16.85$_{-2.06}^{+0.28}$ & 	  3.84$_{-1.46}^{+0.20}$ \\
  D8  & 25 & 		M & 	 68.86 & 	 3.39$_{-0.02}^{+0.02}$ & 	 3.11$_{-0.04}^{+0.04}$ & 	 16.90$_{-0.07}^{+0.07}$ & 	  3.82$_{-0.05}^{+0.05}$ \\
 & & 	K & 	166.94 & 	 6.35$_{-0.09}^{+0.09}$ & 	 2.86$_{-0.10}^{+0.20}$ & 	 16.90$_{-0.19}^{+0.31}$ & 	  3.82$_{-0.13}^{+0.22}$ \\
 & & 	W & 	 84.65 & 	 6.56$_{-0.10}^{+0.19}$ & 	 2.96$_{-0.01}^{+0.19}$ & 	 16.90$_{-0.14}^{+0.38}$ & 	  3.82$_{-0.10}^{+0.27}$ \\
 D10  & 9 & 		M & 	  3.84 & 	 2.40$_{-0.04}^{+0.06}$ & 	 1.61$_{-0.07}^{+0.09}$ & 	 16.39$_{-0.12}^{+0.15}$ & 	  4.02$_{-0.08}^{+0.11}$ \\
 & & 	K & 	 13.08 & 	14.96$_{-3.15}^{+0.04}$ & 	 2.25$_{-0.10}^{+0.19}$ & 	 16.39$_{-4.46}^{+0.28}$ & 	  4.02$_{-3.16}^{+0.20}$ \\
 & & 	W & 	 13.62 & 	14.96$_{-1.85}^{+0.04}$ & 	 2.25$_{-0.10}^{+0.19}$ & 	 16.39$_{-2.63}^{+0.28}$ & 	  4.02$_{-1.86}^{+0.20}$ \\
 D14  & 14 & 		M & 	 19.00 & 	 3.89$_{-0.07}^{+0.06}$ & 	 3.57$_{-0.11}^{+0.09}$ & 	 17.35$_{-0.18}^{+0.16}$ & 	  3.64$_{-0.13}^{+0.11}$ \\
 & & 	K & 	 29.65 & 	 5.20$_{-0.10}^{+0.10}$ & 	 3.21$_{-0.10}^{+0.10}$ & 	 17.35$_{-0.20}^{+0.20}$ & 	  3.64$_{-0.14}^{+0.14}$ \\
 & & 	W & 	 22.75 & 	 5.20$_{-0.14}^{+0.04}$ & 	 3.39$_{-0.19}^{+0.10}$ & 	 17.35$_{-0.33}^{+0.15}$ & 	  3.64$_{-0.24}^{+0.11}$ \\
\hline
\end{tabular}
\hfill\parbox[t]{\textwidth}{Col (1): Cluster name.  Col (2): Number of points used in the fitting procedure.  Col (3): Fitted model, M (Moffat-EFF), (K) Dynamical King, (W) Wilson.  Col (4):  Minimum value of $\chi^2$ obtained for the selected model in Col (3).  Col (5): Shape parameter, $W_{\rm 0}$ for Wilson and Dynamical King models, and $\gamma$ for Moffat-EFF.  Col (6):  Scale parameter, $r_{\rm 0}$ for Wilson and King, and $r_{\rm d}$ for Moffat-EFF.  Col (7--8): Central surface brightness in magnitude and luminosity units, respectively.  The full table is shown in the electronic edition;  a portion is shown here for guidance.} 
\end{table*}

\begin{figure}
\begin{center}
\includegraphics[width=1.0 \columnwidth]{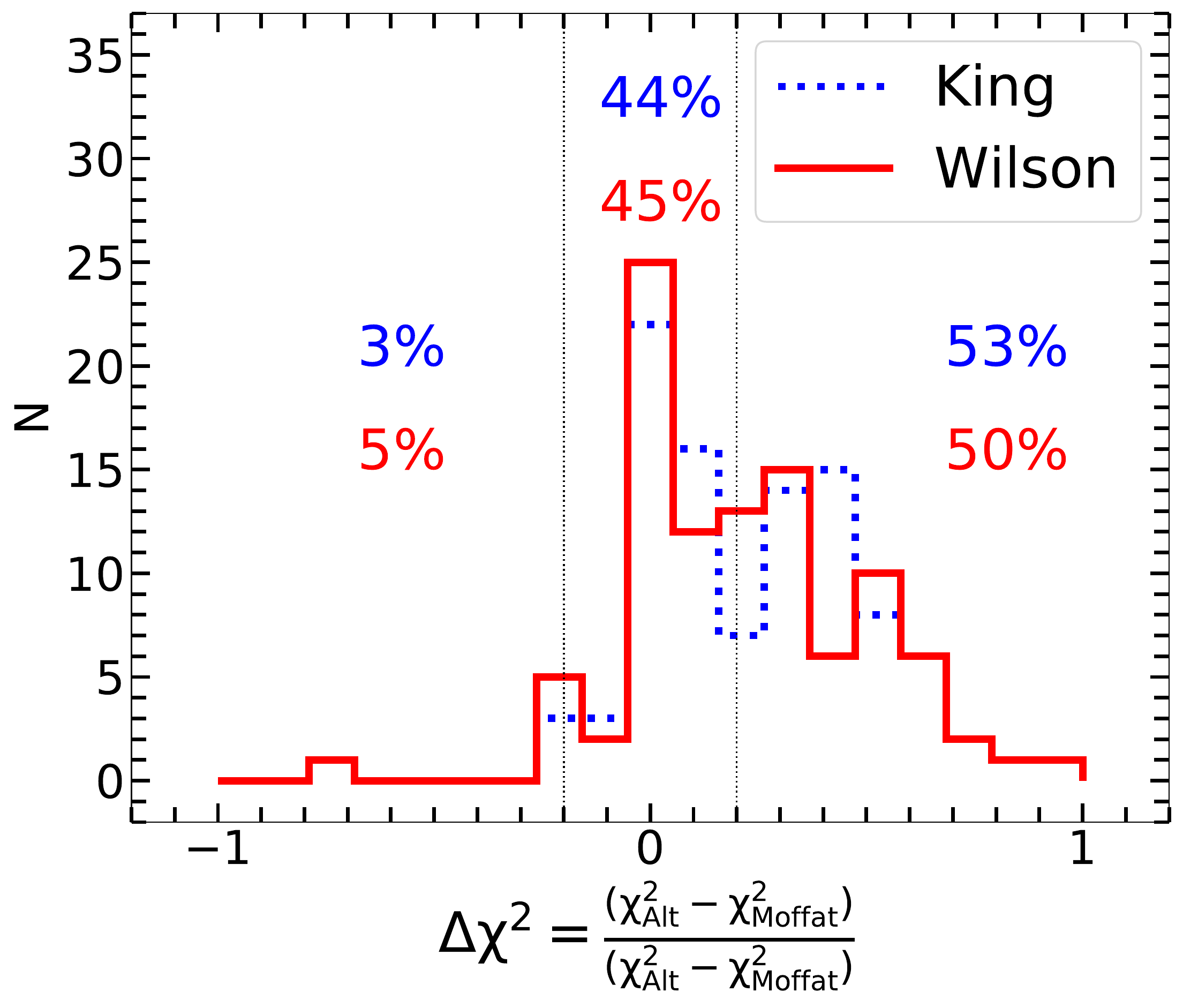}
\caption{$\Delta \chi^2$ distributions of the Moffat-EFF selected as reference model, compared with King (blue dotted) and Wilson (red solid) models for fits in the V-band. The reference and the comparison models are considered to be equally good for $|\Delta \chi^2|\leq$0.2, which is signalled by the vertical dotted lines.  The three columns of numbers appearing in percentage correspond to the King vs Moffat-EFF (blue) and Wilson vs Moffat-EFF (red), for $\Delta \chi^2<-0.2$, $|\Delta \chi^2|\leq$0.2, and  $\Delta \chi^2>0.2$, respectively. For 44--45\% of the clusters, the fits are equally good for any of the three models, with Moffat-EFF model providing a better fit ($\Delta \chi^2>0.2$) for a further 50--53\%.  For only 3-5\% of the clusters ($\Delta \chi^2<-0.2$), King or Wilson models provide a better fit than Moffat-EFF model.  }
\label{fig:delta_chi_all}
\end{center}
\end{figure}

\subsection{Method to select the best model}\label{Subsec:best-fit}

A fit is considered to be good if $\chi^2_{\rm min}$ is of the order of the degrees of freedom ($\nu$) \citep{Wall_statistics}, which in our general case is $\nu=\rm Npts-2$.  However, in fitting SBPs it is common to obtain $\chi^2_{\rm min} < \nu$, even when fits are good \citep{McLaughlin2005}.  This is because, it is necessary to sample the central parts at steps of at least 1 pixel in order to define the SBPs, which is more than a factor of two oversampled with respect to the typical PSF values.  This makes the SBP values at successive points not completely independent of each other, i.e. $\sigma_i$ of successive points are correlated making  $\chi^2_{\rm min} < \nu$. We used the rms errors in the azimuthally averaged intensities as $\sigma_i$ in the $\chi^2$ equation, and hence there may be some contribution to the $\sigma_i$ from real azimuthal variations, which also will make $\chi^2_{\rm min} < \nu$. Some SSCs have $\chi^2_{\rm min} >3 \nu$, which implies the best-fit model does not represent perfectly the observed SBP.

In Fig. \ref{fig:fittings_all}, we show the results for the best fitting parameters for an illustrative cluster, for each of the three model profiles.  In the left-most panel, we show the results for the best-fitting Moffat-EFF model, and in the other two panels, we show the results for the best-fitting Wilson and King models. In the second panel, we show an RGB image for the same cluster. In each panel, fits are shown for the three bands used in this analysis, along with the best-fit model parameters in each band. The \Rfit\ in each band is indicated by vertical lines and \Rfit\ for the V band is shown in the RGB image. \Rfit\  values in B and V bands match in general, whereas in the I band, it is generally smaller.
Bottom panels show the residual $\Delta \mu= \mu_{\rm obs} -\mu_{\rm model}$. 

In most cases, $\chi^2_{\rm min}$ values for the best-fit Moffat-EFF, King and Wilson models are not very different, implying that the fits are equally good for more than one model.  \cite{McLaughlin2005} proposed a method to determine quantitatively the best among these models. We adopted their technique for fits obtained for each filter for every cluster.  This method consists of defining a $\Delta \chi^2$ as:

\begin{equation}
\Delta \chi^2 = \frac{\chi^2_{\rm alt} - \chi^2_{\rm ref}}{\chi^2_{\rm alt} + \chi^2_{\rm ref}},
\label{eq:delta_chi}
\end{equation}
where $\chi^2_{\rm ref}$ and $\chi^2_{\rm alt}$ are the $\chi^2_{\rm min}$ values of the reference model and the model to be compared. Two models are considered to be equally good if $|\Delta\chi^2|\leq$0.2, whereas the reference model is good for $\Delta\chi^2>0.2$.  

In Fig. \ref{fig:delta_chi_all}, we show the $\Delta \chi^2$ distributions for all the 99 SSCs with Moffat-EFF as the reference model.   Around 45\% of the clusters have $|\Delta \chi^2|\leq$0.2, indicating that all the three models are fit equally well for these clusters.   Moffat-EFF models are good fits for 95--97\% of the SSCs, with only 3--5\% of SSCs requiring King or Wilson models. These conclusions also apply to the fits in the other two bands, but with the best-fit Moffat-EFF percentage being around 15\% lower.

\begin{figure}
\begin{center}
\includegraphics[width=1.0 \columnwidth]{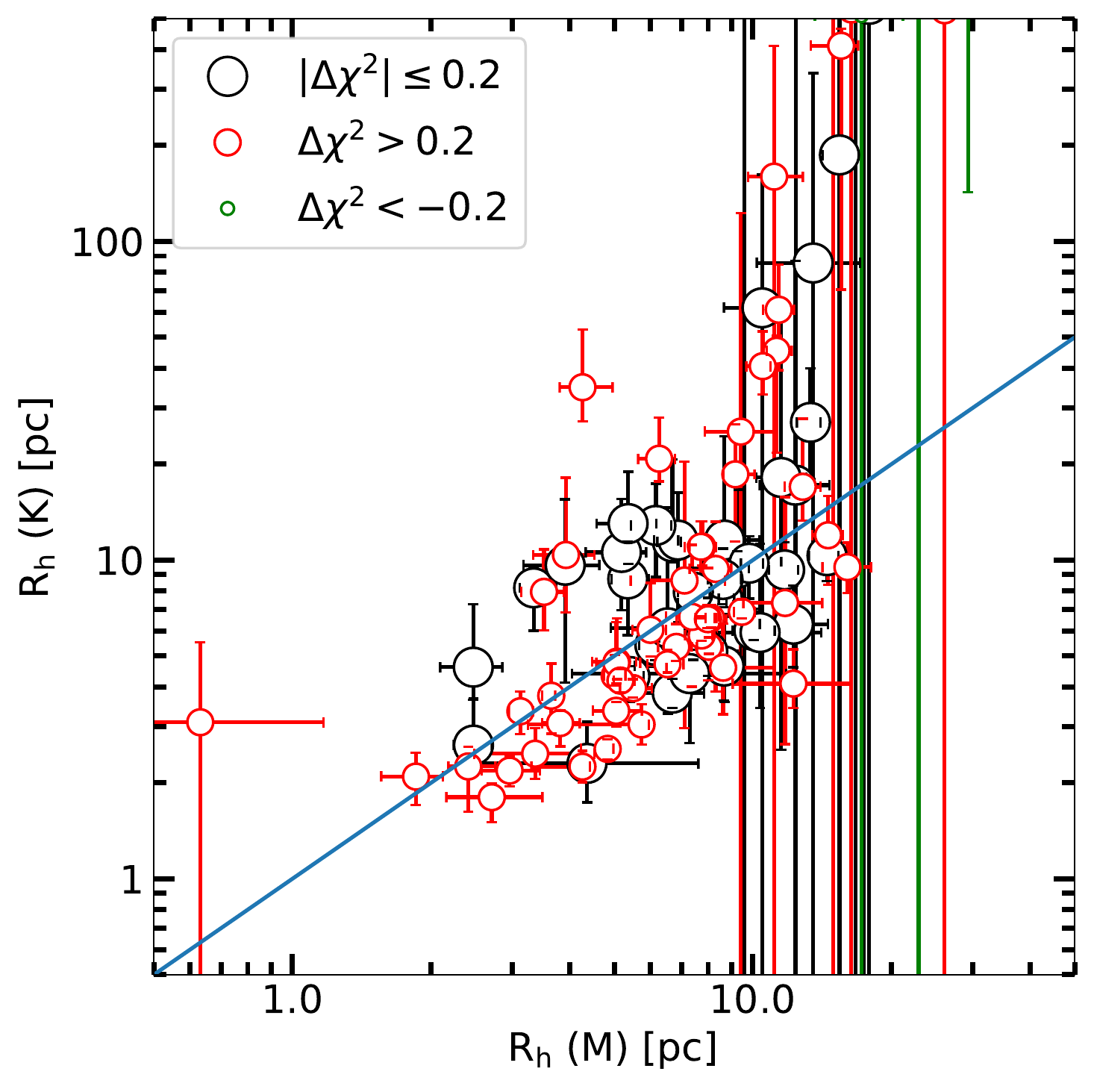}
\caption{Half-light radius $R_{\rm h}$ of King models against those obtained from Moffat-EFF models for the 99 SSCs.  Clusters are colour and size coded to indicate equally good fits with King and Moffat-EFF models (black large circles),  better fit with Moffat-EFF models (red medium-sized circles), and those well represented with King models (green small circles).  The $R_{\rm h}$ derived from both the models agree with each other for $R_{\rm h}\lesssim$10~pc, independent of which model produced the best fit. Beyond this radius, the error bar on the $R_{\rm h}$ derived from King models is systematically larger than those for the Moffat-EFF models.  We note that this is even true for the three clusters for which King model produced the best fit (green vertical lines).}
\label{fig:reff_vs_reff_chi}
\end{center}
\end{figure}

Thus, in general, M82 SSCs are well represented by Moffat-EFF models, and hence we will use the parameters obtained by Moffat-EFF in the V-band as the characteristic values for all clusters.  An examination of the half-light radius $R_{\rm h}$ values indicates that even for the 45\% of the clusters represented equally well by any one of the three models, Moffat-EFF parameters are more reliable than the other two models.  In Fig \ref{fig:reff_vs_reff_chi}, we illustrate this, where we plot the $R_{\rm h}$ of King models against those obtained from Moffat-EFF models for the 99 SSCs.  The error bars on $R_{\rm h}$ are obtained by propagating the errors on the basic derived parameters for each model (See \S\ref{Sec:chi_an_errs}). Clusters for which fits are equally good with King and Moffat-EFF models are distinguished from those for which Moffat-EFF models are good.  It can be seen that $R_{\rm h}$ values for King models are overestimated in several cases independent of if King is a good fit or not.          

\begin{figure}
\begin{center}
\subfloat{\includegraphics[width= \columnwidth]{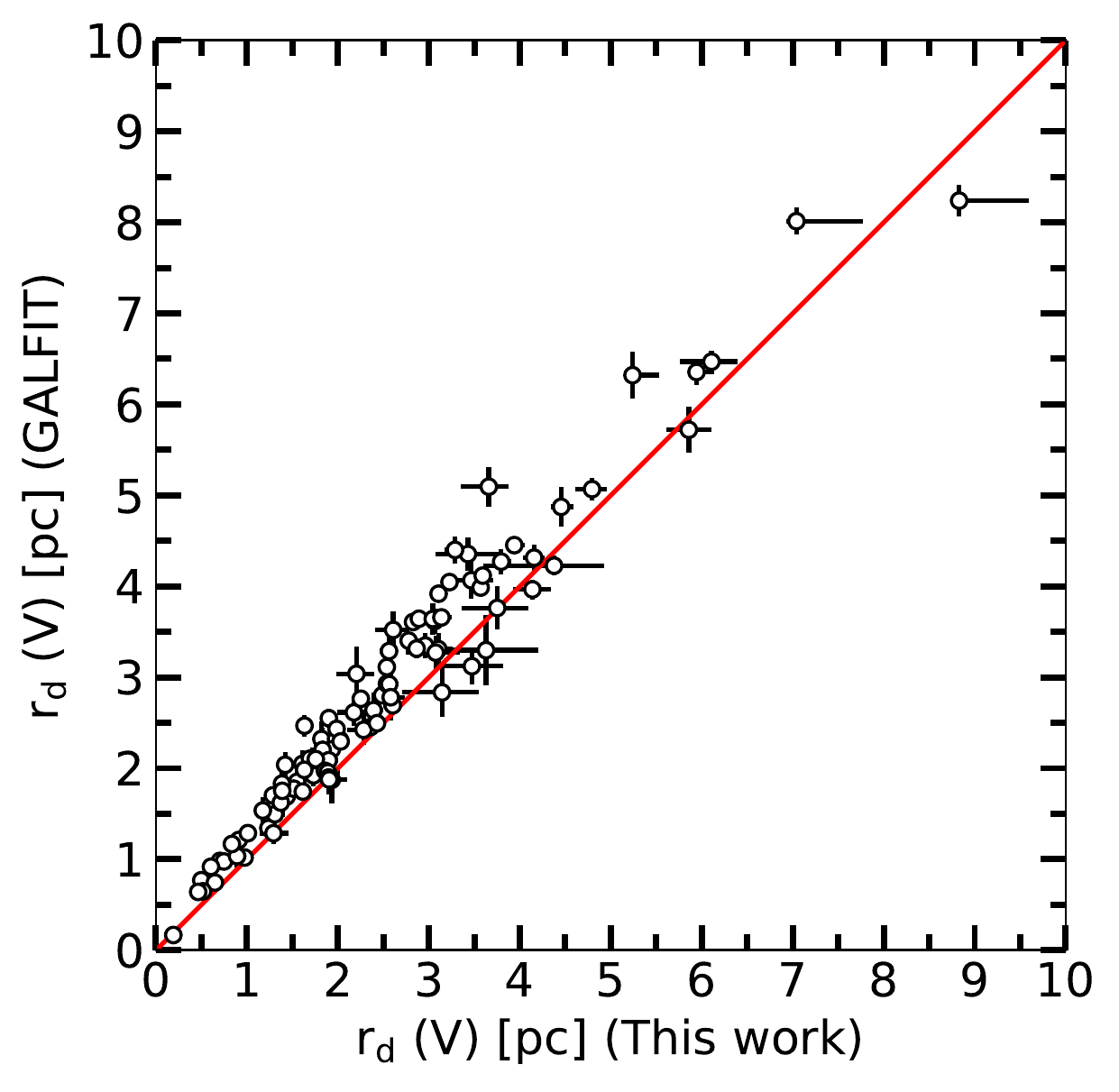}}\\
\subfloat{\includegraphics[width= \columnwidth]{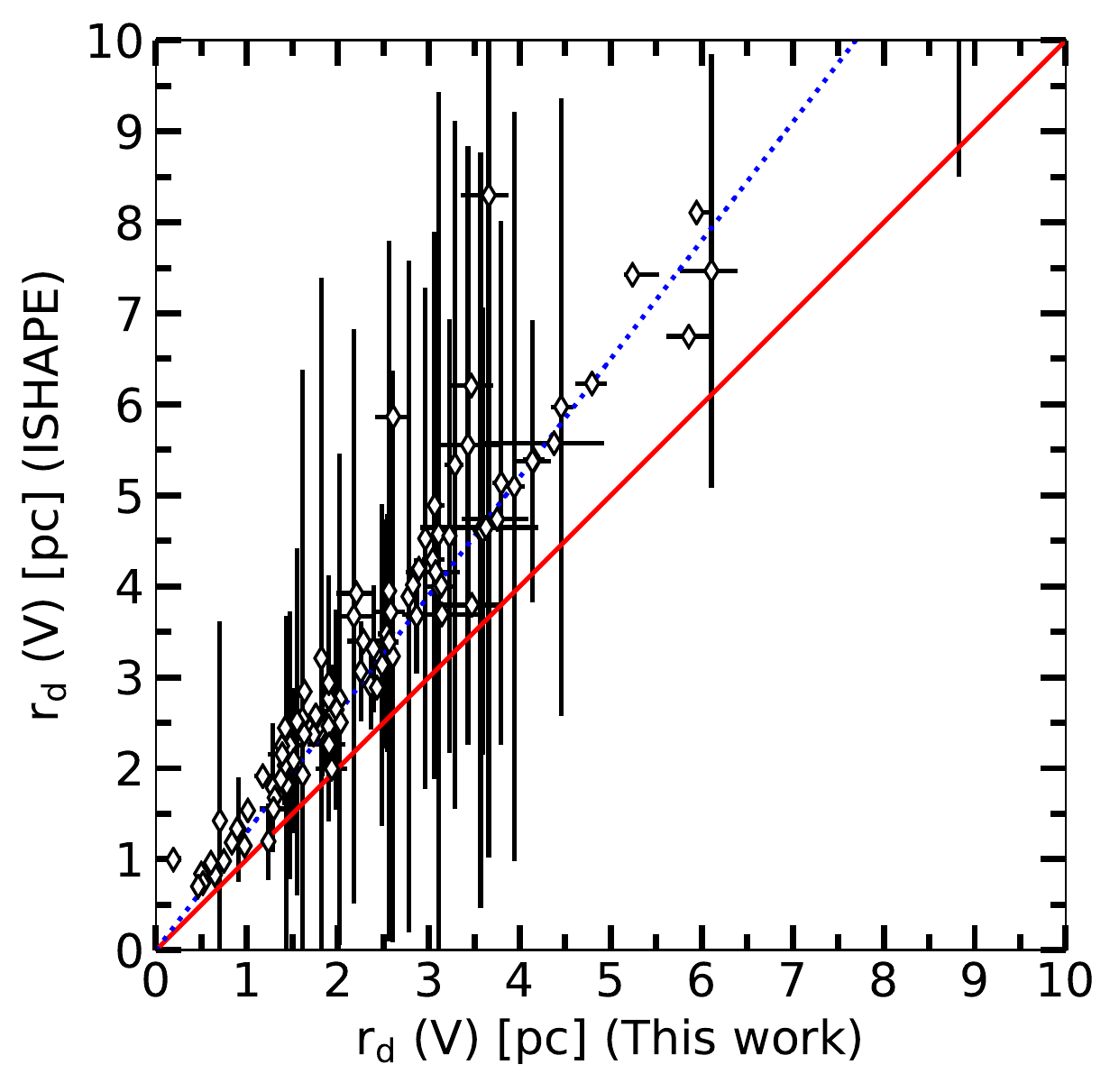}}
\caption{Comparison between Moffat-EFF $r_{\rm d}$ obtained with our own code with that obtained 
using GALFIT (top) and ISHAPE (bottom). The identity function 
is shown with a red solid line. Values from our code are in excellent
agreement with that from GALFIT, whereas the ISHAPE-derived values
are systematically higher by $\sim$30\%, which is indicated by a
blue-dotted line of slope=1.3.}
\label{fig:comp_galfit_ishape}
\end{center}
\end{figure}

\begin{figure*}
\begin{center}
\includegraphics[width= \textwidth]{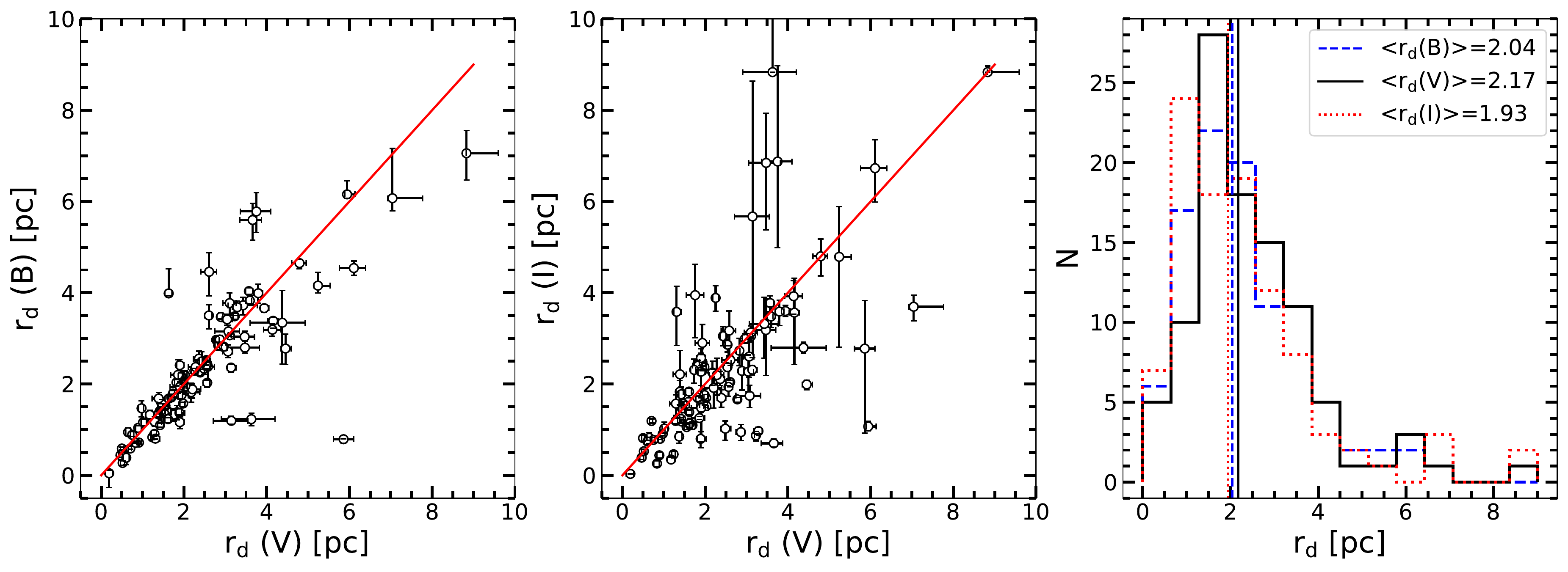}
\caption{Comparison of $r_d$ values  obtained in three filters:  $r_d$ (B) vs $r_d$(V) (left), and $r_d$ (I) vs $r_d$(V) (middle). The red solid line shows the identity function. Histogram of the Moffat-EFF $r_d$ of the 99 selected clusters for the B (blue dashed line), V (black solid line) and I (red dotted line) are shown in the right panel.  The median of each distribution is shown by a vertical line of the same colour code.}
\label{fig:rd_dist}
\end{center}
\end{figure*}

\begin{figure*}
\begin{center}
\includegraphics[width= \textwidth]{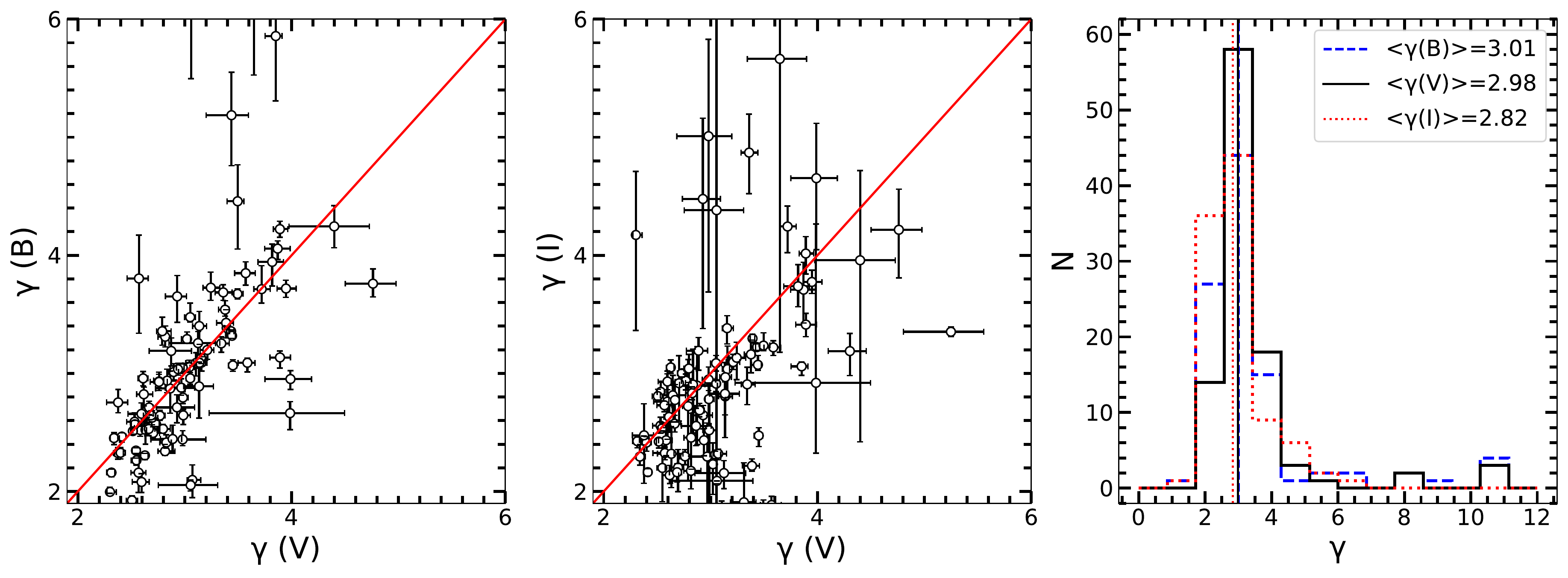}
\caption{Comparison of $\gamma$ values obtained in three filters:  $\gamma$ (B) vs $\gamma$(V) (left), and $\gamma$ (I) vs $\gamma$(V) (middle). The red solid line shows the identity function. Histogram of the Moffat-EFF $\gamma$ of the 99 selected clusters for the B (blue dashed line), V (black solid line) and I (red dotted line) are shown in the right panel.  The median of each distribution is shown by a vertical line of the same colour code.}
\label{fig:beta_dist}
\end{center}
\end{figure*}

\begin{figure*}
\begin{center}
\includegraphics[width= \textwidth]{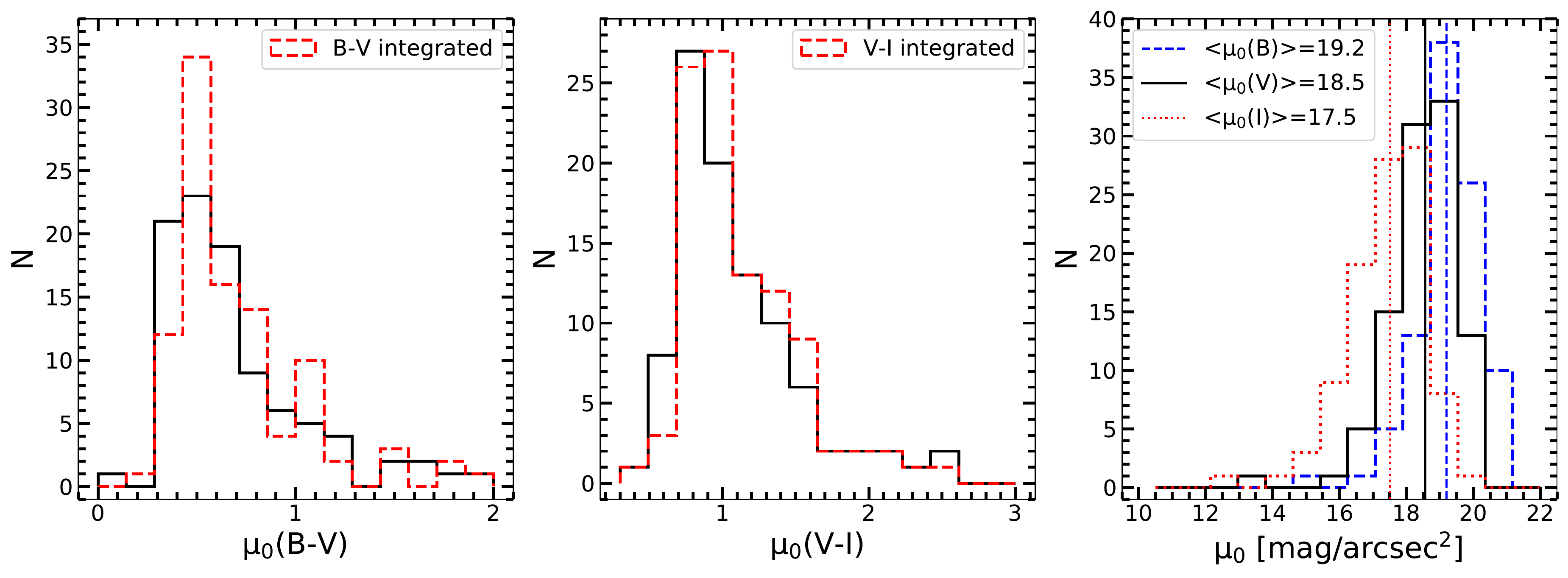}
\caption{Central (black solid line) and integrated (red dashed line) colour histograms for $\mu_0(B)-\mu_0(V)$ (left), and $\mu_0(V)-\mu_0(I)$ (middle).  Histogram of the Moffat-EFF $\mu_0$ of the 99 selected clusters for the B (blue dashed line), V (black solid line) and I (red dotted line) are shown in the right panel.  The median of each distribution is shown by a vertical line of the same colour code. }
\label{fig:muo_dist_sigma}
\end{center}
\end{figure*}

In Table \ref{tab:basic_fit_pars}, we show the best-fitted shape ($W_0$ or $\gamma$), scale ($r_0$ or $r_d$) and central surface brightness ($\mu_0$ and $I_0$) parameters for all the three models along with the $\chi^2_{\rm min}$ values for the fits in the V-band.   
Data for each cluster are organised in three rows:  the first row shows the results for Moffat-EFF, the second row for King models and the third row for Wilson models.  
For the last four columns, we give their respective error bars.  The error bars on shape and scale parameters are based on the analysis in \S\ref{Sec:chi_an_errs}.  The error bars in the central surface brightness are based on these errors propagated in quadrature, following the prescription of \citet{McLaughlin2005}.   
The $\chi^2_{\rm min}$ given in Column 4 is related to the reduced $\chi^2_{\nu}$ by the number of degrees of freedom $\nu$, which in our case is equal to Npts$-2$ given that we have two fitted parameters.  Hence, $\chi_\nu^2=\frac{\chi_{\rm min}^2}{\rm Npts-2}$.  For all of our clusters, $\chi_\nu^2<$3 in the V-band for at least one of the three models. 

\subsection{Comparison with GALFIT and ISHAPE}\label{Sec:comp_galf}

GALFIT \citep{Penggalfit} and ISHAPE \citep{Larsenishape} are two widely used tools 
for obtaining structural parameters of extragalactic clusters. In order to 
ensure that there are no systematic offsets
in the values of structural parameters obtained by our fitting tool 
with those obtained with these two tools, we carried out the
fits with the Moffat-EFF profiles on all our sample clusters with these two tools.
The PSF images, as well as the fitting radius for each cluster are retained 
from our analysis. In the case of ISHAPE, we oversampled our PSF image 
by a factor of ten using the tool \emph{magnify} in IRAF, as required by the code.
Fittings were carried out on 2-D images of 101$\times$101 pixel cut-outs.
Both these codes have their own algorithm for background determination.
The $\gamma$ values are left free for GALFIT, whereas for ISHAPE, we used 
our best-fit $\gamma$ values for each cluster.
In Fig. \ref{fig:comp_galfit_ishape}, we compare our $r_{\rm d}$ 
values with those from GALFIT and ISHAPE.
We observe that in general, our values are in excellent agreement with those of GALFIT,
and are consistent within the errors with those of ISHAPE, but with a slope
of 1.3, instead of unity.
The lengths of the error bars in GALFIT are very similar to ours, whereas
ISHAPE values have larger error bars. We checked that these error bars and the
values do not vary much for a fixed value of $\gamma=$3.
 
\section{Results and discussions}\label{Sec:results}

In the previous section, we concluded that Moffat-EFF models adequately represent 
all our subsample of 99 SSCs.  Model-fitting directly gives us four parameters, 
namely $r_{\rm d}$, $\gamma$, $\mu_{\rm 0}$ and $L_{\rm tot}$. Core radius $R_{\rm c}$, 
the radius at which the surface brightness is half its peak value, is related
to $r_{\rm d}$ through Eqn.~\ref{rc_rd}. 
The half-light radius $R_{\rm h}$, is another parameter that can be calculated 
from these parameters (see \S\ref{Subsec:best-fit}). 
Dynamical analysis of the fitted model profiles, along with a knowledge of mass-to-light ratio for M82 SSCs, allows us to calculate four more parameters: mass, the central velocity dispersion $\sigma_0$, central mass density $\rho_0$ and central mass surface density $\Sigma_0$. 
Not all these parameters are independent of each other. 
Evolved objects like GCs show a tight inverse correlation between $R_{\rm c}$ and $\mu_{\rm 0}$ \citep{Kormendy1985}.
These correlations are part of the Fundamental Plane for GCs \citep{Djorgovski1995,McLaughlin2000}. A detailed analysis of all the derived parameters will be carried out in a forthcoming paper.  In this paper, we will characterise the basic parameters obtained in the three bands.

\subsection{Colour-dependence of the derived parameters}

We have carried out independent analysis of SBPs in three filters for all our sample SSCs to study the possible colour-dependence of the derived parameters.
In Figs. \ref{fig:rd_dist}, \ref{fig:beta_dist}, and \ref{fig:muo_dist_sigma}, we compare the distribution of $r_{\rm d}$, $\gamma$, and $\mu_{\rm 0}$ for Moffat-EFF models in the three filters.  
The intention of showing these distributions is to compare the shape, as well as the center of the distributions.
Median values are indicated in all the plots by vertical dashed lines.

Median values of $r_{\rm d}$ in the three filters are very similar with a value $\sim$2.0~pc.  
In all the three filters, the distribution is asymmetric with its peak lying at $\sim$1~pc to the left of the median value, and having a long tail that reaches up to $\sim$7--9~pc. 
The behaviour of $\gamma$ distribution is very similar to that of $r_{\rm d}$ with the median value of $\gamma\sim3.0$ for the three filters.  $\mu_{\rm 0}$ median values in the three filters are:  $\mu_{\rm 0}$(B)=19.2, $\mu_{\rm 0}$(V)=18.5, $\mu_{\rm 0}$(I)=17.5 mag\,arcsec$^{-2}$.
From these plots, we conclude that the distributions in the three filters are similar.   Hence, we use the values in the V band in the rest of our analysis.

\subsection{Functional form of $\gamma$ distribution}

Following the seminal study of \citet{Elson}, power-law form of the SBPs represented 
by the Moffat-EFF profile, is considered to be the characteristic feature of 
young SSCs.  On the other hand, King profiles \citep{King_emp} are applicable to more evolved systems such as GCs. 
\citet{Elson} found the power-law form extends to beyond the tidal radius in young
clusters. They argue that clusters take around 2 to 3 orbital periods
to get rid of the stars outside the tidal radius, and hence, have to be older than $\sim$1~Gyr to show a King SBP. 
Until that time, the escaped stars would be located in an unbound halo.

\citet{MackeyGilmore2003a} analysed the SBPs obtained from HST images of a 
sample of 53 LMC clusters using Moffat-EFF profiles. 
With as much as 25 clusters in this sample being younger than $\sim$1~Gyr, 
this happens to be the only case where young and intermediate-age well-resolved clusters have been analysed using a uniform set of Moffat-EFF parameters.
For this reason, the parameter set obtained by \citet{MackeyGilmore2003a}
has become the benchmark against which parameters of SSCs in other
galaxies have been compared with.
Our analysis of nearly 100 intermediate-age (50--300~Myr) SSCs,
offers an opportunity to understand the transition
from power-law shaped young clusters to King profile shaped GCs.

\begin{figure}
\begin{center}
\subfloat{\includegraphics[width=.45\textwidth]{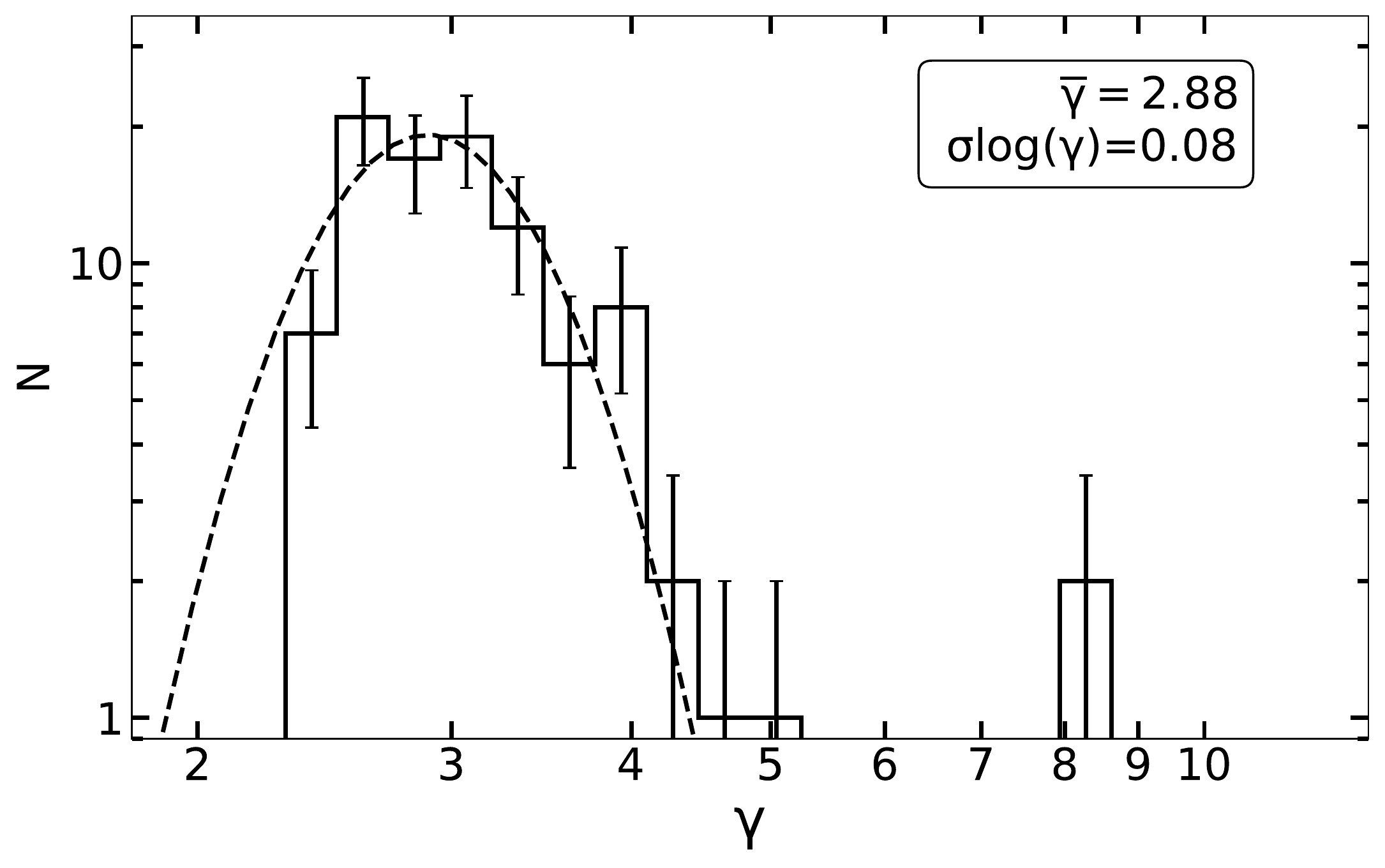}}\\
\subfloat{\includegraphics[width=.45\textwidth]{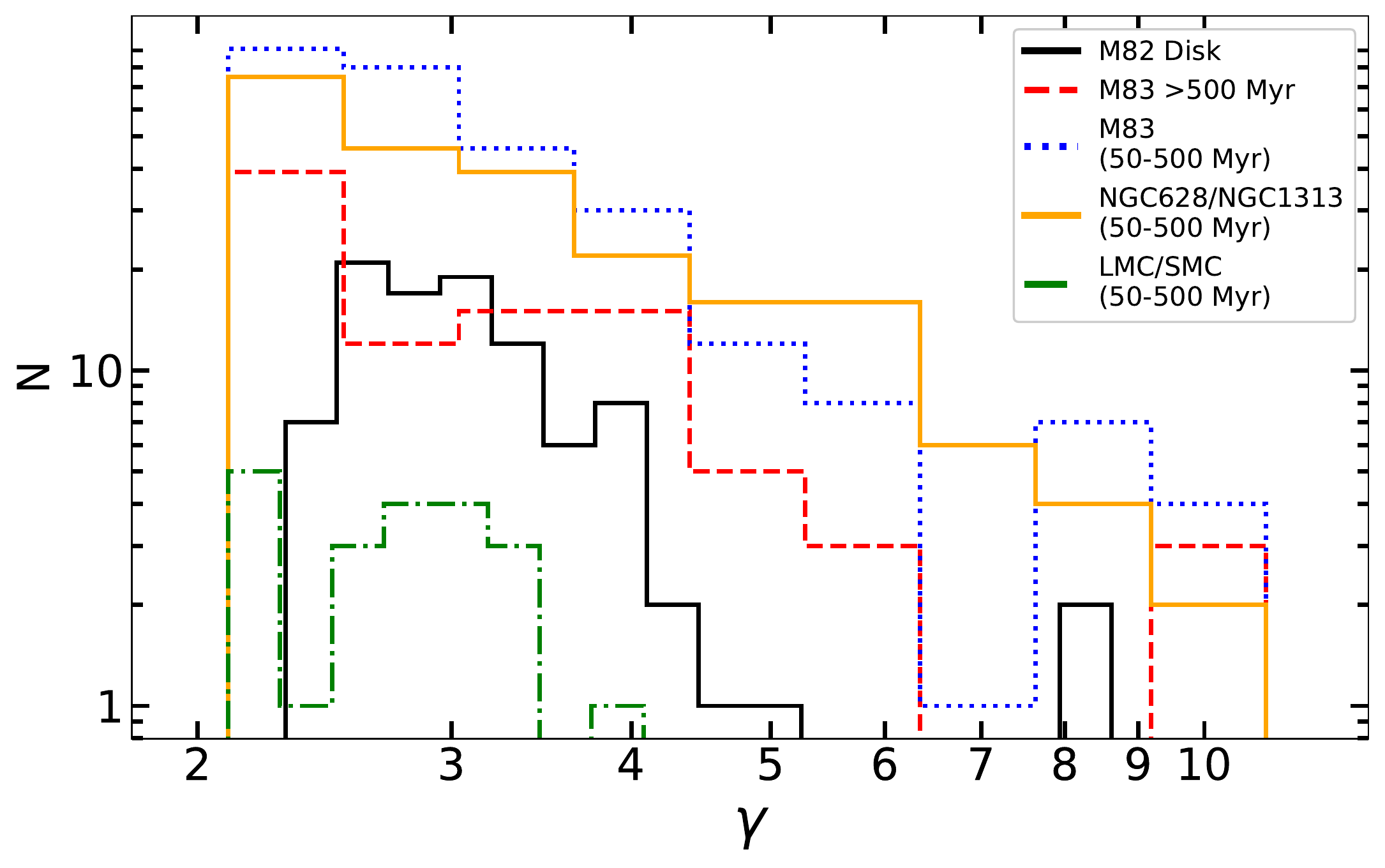}}
\caption{Upper panel: distribution of $\gamma$ parameter for 99 M82 SSCs fitted with Moffat-EFF profile (histogram).  The error bars are based on Poisson statistics.  In dashed line, we show a log-normal fitting with a peak value of  2.88 and a standard deviation of $\rm \sigma log(\gamma)=0.08$.  Bottom panel: distribution of $\gamma$ parameter in other nearby galaxies are compared with that of M82.}
\label{fig:dist_beta_sigma}
\end{center}
\end{figure}

The $\gamma$ measures the slope of the power-law SBP at large radii (see Eqn. \ref{eq:moffat_proj}).
A $\gamma$=2 corresponds to the case of a King profile with an infinite value of
concentration parameter, and infinite mass. For real clusters, $\gamma>$ 2. The higher
the value of $\gamma$, the steeper is the outer slope. 
In Fig \ref{fig:dist_beta_sigma} (top panel), we show the distribution of $\gamma$ for M82 SSCs.
The distribution is well represented by a log-normal function of $\rm \sigma log(\gamma)=0.08$, 
centered at $\gamma$=2.88, which is close to the median value of 3.0.
This value agrees well with the median value found for clusters in the LMC and other nearby
galaxies \citep{Portegiesrev}.  
In the bottom panel, we compare the distribution of $\gamma$ for the disk clusters in M82,
with those in other galaxies (LMC/SMC, M83, NGC1313 and NGC628) 
where measurements of $\gamma$ had been carried out.
Parameters for M83 come from the study of \citet{Ryon} and NGC1313 and NGC628
from \citet{Ryon2017}. These parameters were obtained using GALFIT.
LMC/SMC cluster parameters come from the study of \citet{MackeyGilmore2003a, MackeyGilmore2003b}.
We divided the sample in these galaxies into young ($<$50~Myr), intermediate-age (50--500~Myr) and old ($>$500~Myr) clusters.  For the sake of comparison
with M82 disk SSCs, we use the sample of intermediate-age clusters.
The subsample of intermediate-age clusters  includes 335 in M83, 
235 in NGC628, 147 in NGC1313, and 24 in LMC/SMC. 
 
Our distribution compares well with that in the LMC/SMC, both being log-normal
centered around $\gamma=$2.9. On the other hand, $\gamma$-values distribute over
a wide range in other galaxies, peaking at the minimum value of $\gamma$=2, and 
decreasing almost linearly (power-law) on this plot for higher values.
For M83, we show the distribution of old clusters also.
The sample of intermediate-age clusters of M83 shows the same behaviour as for
the old sample, and hence the power-law tendency seems to be independent of 
evolutionary stage.

\citet{Elson} argued that $2.5<\gamma<3.2$ correspond to density profiles in dissipationless
systems. Cluster formation in their parent molecular cloud should be 100\% efficient
for the real clusters to be dissipationless. 
On the other hand, real clusters are expected to contain some residual gas within 
the cluster volume, which would be expelled from the cluster
in the first 10~Myr, when massive stars end their lives as supernovae. 
The loss of gravitational energy of the expelled gas-mass makes the cluster 
expand, which eventually shapes the outer part of the density profile 
\citep{BastianGoodwin2006}.
As the residual gas fraction
or equivalently efficiency of cluster formation is expected to vary from one cluster to
the other, real clusters are expected to have a wide range of $\gamma$ values,
in this scenario of cluster formation.
An alternative scenario is that the gas continues to flow into the cluster volume
even after the first supernovae explosions \citep{Fujii2012,Parker2014}. 
This is recently found to be the case in dense progenitor clouds of 
massive clusters in the Milky Way \citep{Walker2015}.
Under this scenario, clusters do not necessarily expand freely following the 
multiple-supernovae explosions \citep{Silich2017}. 
It is likely that such clusters conserve their initial profile shape.

The log-normal form and the small spread in the $\gamma$ value 
seem to support the latter scenario of cluster formation.

\subsection{Functional form of $R_{\rm c}$ distribution}

We now discuss the distribution of $R_{\rm c}$ for our sample of SSCs. With $R_{\rm c}$=0.1~pc \citep{Elson1992,MackeyGilmore2003a} R136 in the LMC, often considered as the prototype for a young SSC,
is one of the most compact SSCs known. Several young extragalactic SSCs also are
found to have sub-parsec values of $R_{\rm c}$ \citep{Portegiesrev}.
\cite{BastianGieles2008} found a tendency for $R_{\rm c}$ to increase with age, which they
interpreted as an evidence for expansion of SSCs.

\begin{figure}
\begin{center}
\subfloat{\includegraphics[width=.45\textwidth]{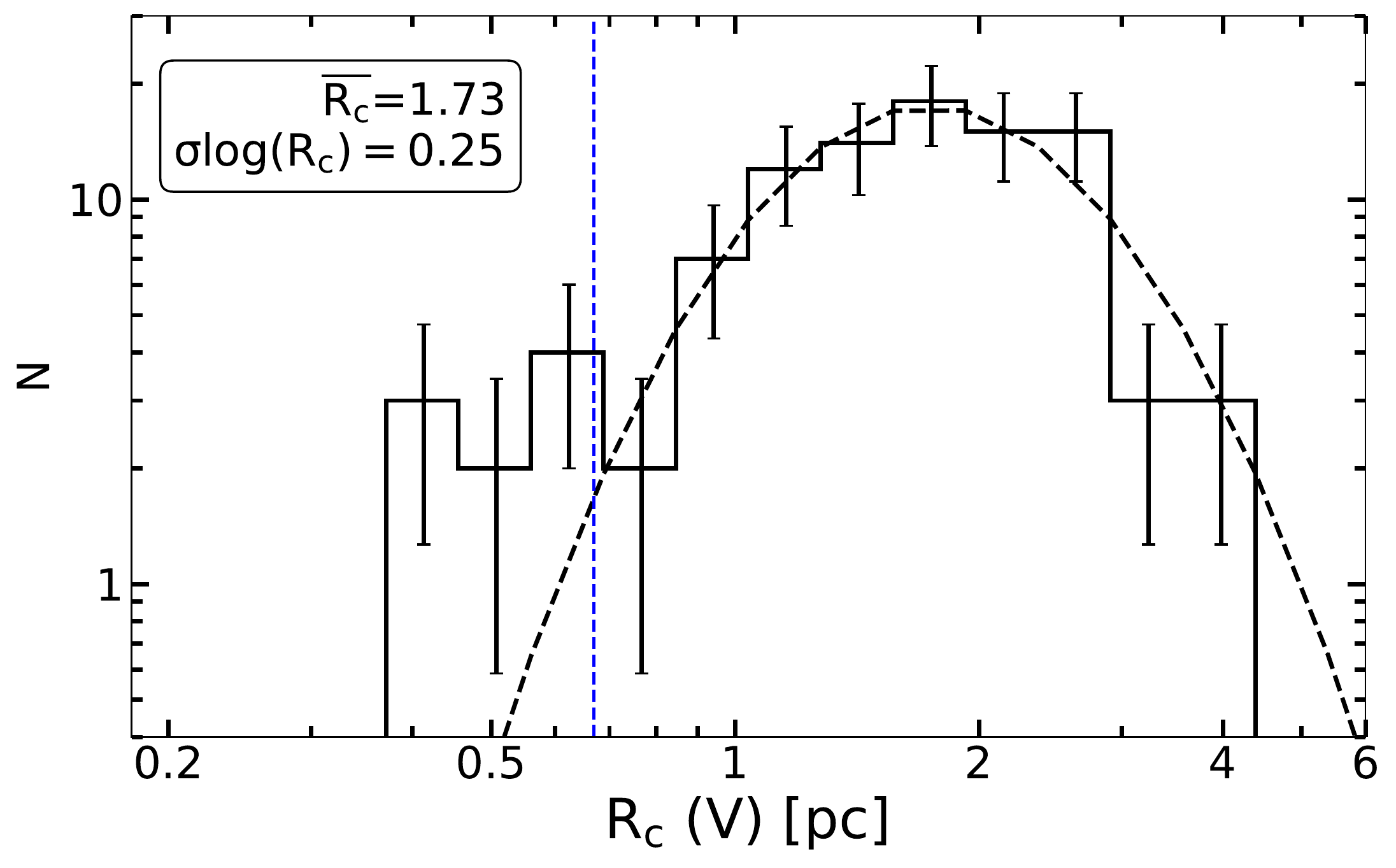}}\\
\subfloat{\includegraphics[width=.45\textwidth]{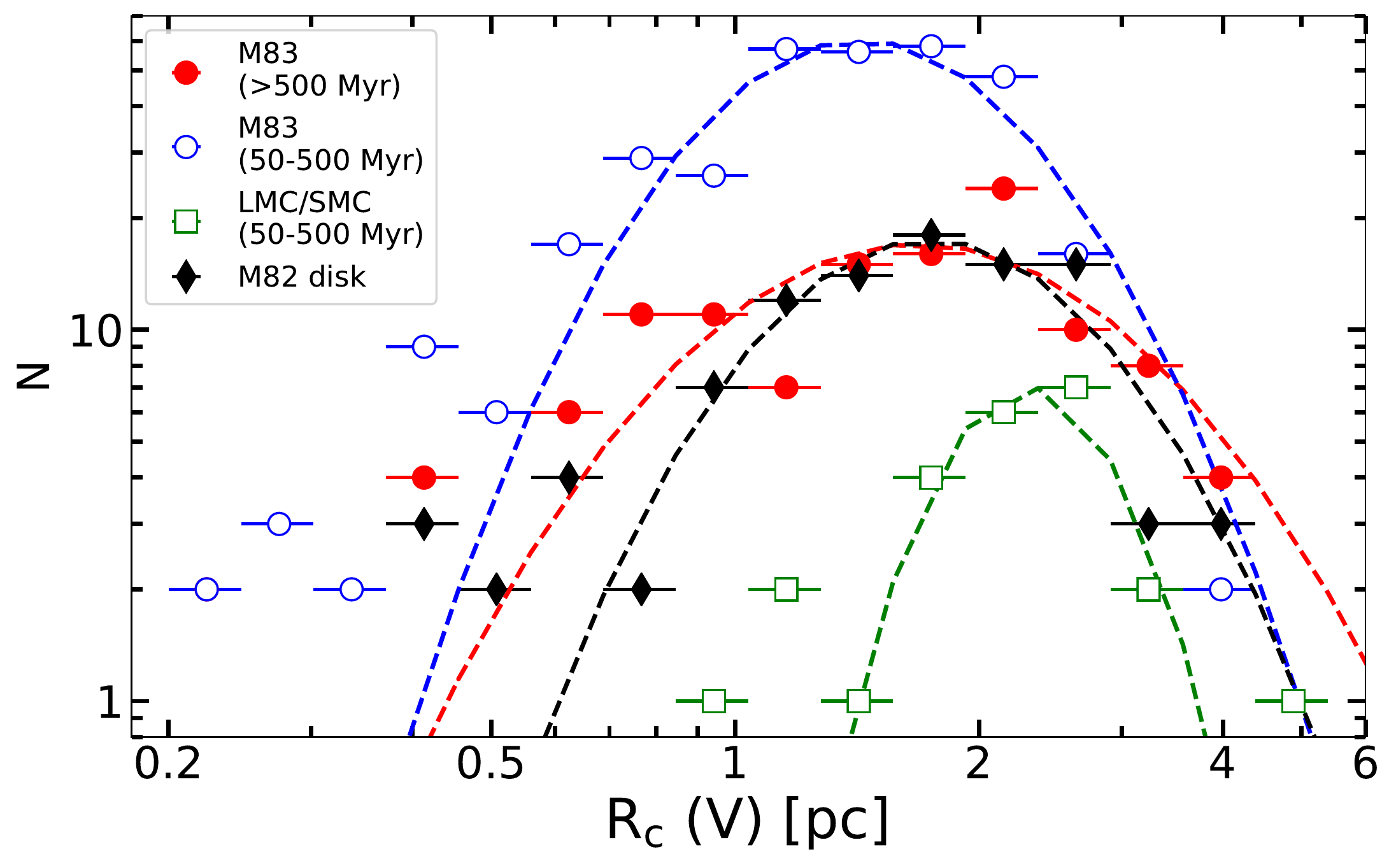}}
\caption{Upper panel: distribution of core radius $R_{\rm c}$ for 99 M82 SSCs fitted with Moffat-EFF profile (histogram).  The error bars are based on Poisson statistics.  In dashed line, we show a log-normal fit to the data, whose parameters are given in the top-left corner.  The minimum reliable value according to the PSF is shown with a vertical blue dashed line.   Bottom panel: comparison of binned distributions (symbols explained in the top-left corner) of $R_{\rm c}$ for M83 old (red), M83 intermediate-age (blue), LMC/SMC (green), M82 (black) SSCs. The horizontal bars correspond to the fixed logarithmic width used for binning. The best-fit log-normal function is shown by dashed lines following the same colour code as the binned data.}
\label{fig:rc_dist_sigma}
\end{center}
\end{figure}

In Fig. \ref{fig:rc_dist_sigma} (top panel), 
we show the distribution of  the core
radius $R_{\rm c}$ from our study. The distribution fits very well with a log-normal
function centered at  
$R_{\rm c}=$1.73~pc, and $\rm \sigma log \big{(} \frac{R_{\rm c}}{pc}\big{)}=0.25$. 
The median value of  the distribution is 1.62~pc, which is close to the peak 
of the log-normal distribution. 
The study of \cite{BastianGieles2008} includes M82 disk SSCs from the spectroscopic sample
of \citet{Konst2009}, for which they report a median value of $R_{\rm c}$=2.2~pc,
which falls well within the range of our $R_{\rm c}$ values.

We used the same dataset as for $\gamma$ to compare our $R_{\rm c}$ values with that in
other galaxies. In the bottom panel, we show a plot comparing the distributions, where   
each distribution is fitted with a log-normal function.  The central value ($\overline{R_{\rm c}}$) and $\sigma$ of the function are given in Table \ref{tab:stat_details}.  LMC and M83 clusters with ages similar to that in M82 disk (50--500~Myr) have mean $R_{\rm c}$ value higher and lower, respectively, as compared to that in M82. Incidentally, the morphological type of these galaxies change from SABc in M83 \citep{rc31991}, Irr II/SBd  in M82 \citep{Mayya2005}, to SBm in LMC \citep{rc31991}.  The progressive increase in mean $R_{\rm c}$ for similar-age clusters is probably suggesting that the morphological type has a role in core evolution at intermediate-ages. 
We also note that at older ($>$500 Myr) ages, M83 clusters have mean $R_{\rm c}$ similar to intermediate-age clusters of M82. This tendency of $R_{\rm c}$  increasing with age in M83 has been reported by \citet{Ryon}.  Similar tendency is also seen in LMC/SMC, which has been attributed to cluster expansion by \citet{Mackey2008}.      

\begin{table}
\begin{center}
\caption{Statistical properties of the Core Radius distributions in M82 and other nearby galaxies.}
\label{tab:stat_details}
\begin{tabular}{lccc}
\hline
 & $\overline{R_{\rm c}}$ & $\sigma\log\big{(} \frac{R_{\rm c}}{pc}\big{)}$ & N\\
(1) & (2) & (3) & (4)\\ 
\hline
M82 (Disk) & 1.73 & 0.25 & 99\\
M83 (50--500 Myr) &  1.42 & 0.51 & 335\\
M83 (>500 Myr) & 1.66 & 0.56 & 118\\
LMC/SMC (50--500 Myr) & 2.29 & 0.39 & 24\\
\hline
\end{tabular}
\end{center}
\hfill\parbox[t]{\columnwidth}{Col (1): Galaxy name and age range.  Col (2): Peak value of $R_{\rm c}$.  Col (3): Standard deviation of the log-normal distribution. Col (4): Number of clusters in the specified age range.} 
 \end{table}

\subsection{$\gamma$ vs $R_{\rm c}$ relation}

Dynamical evolutionary models of clusters by \citet{Mackey2008} find a steady 
increase of both $\gamma$ and $R_{\rm c}$ with age. Different physical processes 
are at work at different time scales. After the early steep increase in radius 
driven by residual gas expulsion, the mass-loss during stellar evolution is 
the principal process that drives the evolution of $\gamma$ and $R_{\rm c}$ 
up to around 600~Myr. Such an evolution of $\gamma$ and $R_{\rm c}$ are observed 
in the clusters in LMC and SMC \citep{MackeyGilmore2003a,MackeyGilmore2003b,Mackey2008}.   

\begin{figure}
\begin{center}
\includegraphics[width= \columnwidth]{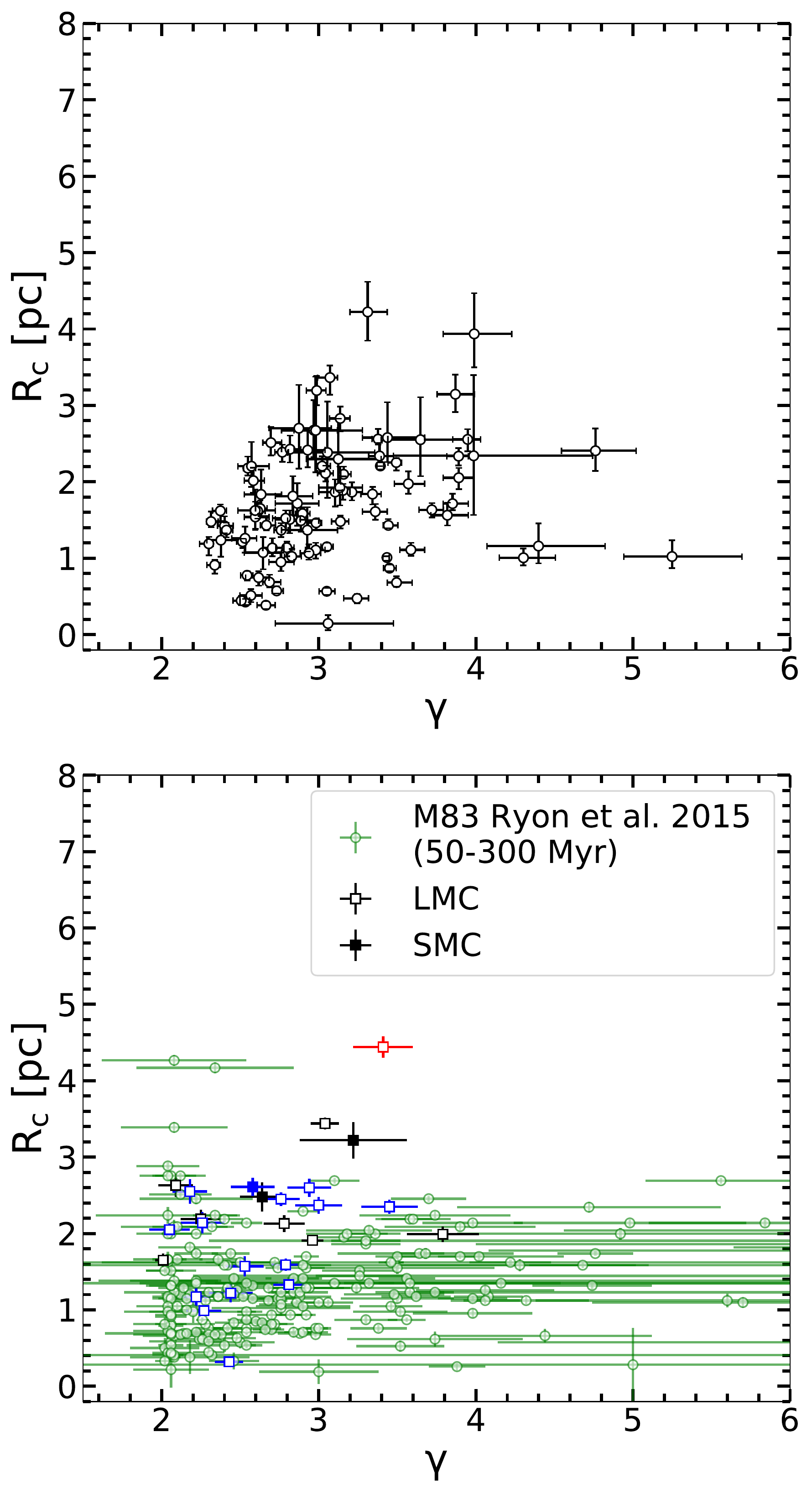}
\caption{$R_{\rm c}$ vs $\gamma$ diagram for the 99  M82 SSCs (top), and the 
clusters in LMC and SMC from \citet{MackeyGilmore2003a,MackeyGilmore2003b}, as 
well as M83 from \citet{Ryon} (bottom), all fitted with Moffat-EFF profiles.   
The bars represent the errors in the $R_{\rm c}$ vs $\gamma$ plane. 
In both the plots, intermediate-age clusters (50--300 Myr) are shown with black 
symbols, and younger and older clusters are shown with blue and red symbols, 
respectively.  The M83 data is shown in a single colour, since it only corresponds 
to the intermediate age range (50--300 Myr).}
\label{fig:beta_vs_rc}
\end{center}
\end{figure}

In Figure \ref{fig:beta_vs_rc}, we plot $R_{\rm c}$ against $\gamma$ for our 
sample of SSCs (top), as well as for SMC and LMC clusters from 
\citet{MackeyGilmore2003a,MackeyGilmore2003b} and M83 from \citet{Ryon} (bottom). 
For the LMC/SMC clusters, a clear trend of the upper envelope of
$R_{\rm c}$ increasing with increasing $\gamma$, as expected in the models of 
\citet{Mackey2008}, is seen. For the M82 sample, the trend is weaker. 
But the trend is also weaker for LMC/SMC and M83 SSCs that have a similar range of ages 
as that of M82 SSCs. Hence, the observed values of  $R_{\rm c}$ and $\gamma$ in M82 are in broad agreement with the predictions of \citet{Mackey2008}. 

\section{Summary}\label{Sec:summary}

In this work, we have carried out structural analysis of 99 intermediate-age (50-300 Myr) 
SSCs in the disk of 
M82 using the intensity profiles derived from HST images in F435W, F555W and 
F814W bands. These clusters have a narrow range of ages
between 50--300~Myr, which provides an excellent opportunity to understand the 
structural parameters at ages intermediate between Young Super Star Clusters 
and Old Globular Clusters. Structural parameters were derived for the King, 
Wilson and Moffat-EFF models, using the standard $\chi^2$ minimisation technique. 
Errors on the extracted parameters were determined based on the $\chi^2$ 
statistics of the fitting models. Experiments on mock clusters were also
carried out to authenticate the extracted parameters as well as their errors.
In order to further validate our fitting technique, we 
obtained structural parameters with the Moffat-EFF profiles for our entire 
Sample of clusters using
widely-used tools such as GALFIT and ISHAPE. We find excellent agreement with
the values and their errors obtained by GALFIT, whereas ISHAPE values have
systematically large errors.

The observed profiles are in general well-fitted by all the three model profiles. 
Using quantitative criteria for discrimination between the models used, we 
find the majority of clusters ($\sim$95\%) is well-represented by Moffat-EFF profiles. 
We tabulate the fitted parameters in the F555W band for all the clusters using the three models, 
and analyse in detail the statistical properties of Moffat-EFF parameters.  
The distributions of $\gamma$ and $r_{\rm d}$ in the three bands are similar, 
with very similar median values. 
The distribution of $\gamma$ follows a log-normal shape around a central value of 
2.88 and $\rm \sigma log(\gamma)=0.08$.  Values of $\gamma<$3 imply the existence 
of an extended halo in M82 clusters. 
The $R_{c}$ distribution also follows a log-normal form with peak values of $R_{c}$=1.73 pc,
and $\rm \sigma log \big{(} \frac{R_{\rm c}}{pc}\big{)}=0.25$.
These values are large as compared to both Young SSCs and Old GCs, but compare 
well with the corresponding values for LMC intermediate-age clusters. 
Our $\gamma$ and $R_{\rm c}$ distributions are also 
compared with the intermediate-age SSCs in M83, NGC1313 and NGC628. 
We find a larger spread of $\gamma$ values in these galaxies as compared to
our log-normal distribution in M82. On the other hand, $R_{\rm c}$ distributions 
in M83 and M82 are comparable, with systematically larger core sizes for M82
SSCs. Detailed analysis of these differences, taking into account cluster masses,
ages, host galaxy properties will be addressed in a forthcoming paper. 

\section*{Acknowledgements}

We greatly appreciate the valuable suggestions we received from an
anonymous referee,
which has improved this manuscript, especially in the presentation of
the last 3 figures.
BCO thanks CONACyT for granting PhD research fellowship that enabled
her to carry out the work presented here. We also thank CONACyT for
the research grants CB-A1-S-25070 (YDM), CB-2014-240426 (IP), and
CB-A1-S-22784 (DRG), that allowed the acquisition of a cluster that
was used for computations in this work.
Special thanks to Sergiy Silich for accepting BCO as his research
assistant during the later stages of this work.

\addcontentsline{toc}{section}{Acknowledgements}       
\bsp	
\label{lastpage}
\bibliographystyle{mnras}
\bibliography{bibliografia}

\begin{thebibliography}{}
\makeatletter
\relax
\def\mn@urlcharsother{\let\do\@makeother \do\$\do\&\do\#\do\^\do\_\do\%\do\~}
\def\mn@doi{\begingroup\mn@urlcharsother \@ifnextchar [ {\mn@doi@}
  {\mn@doi@[]}}
\def\mn@doi@[#1]#2{\def\@tempa{#1}\ifx\@tempa\@empty \href
  {http://dx.doi.org/#2} {doi:#2}\else \href {http://dx.doi.org/#2} {#1}\fi
  \endgroup}
\def\mn@eprint#1#2{\mn@eprint@#1:#2::\@nil}
\def\mn@eprint@arXiv#1{\href {http://arxiv.org/abs/#1} {{\tt arXiv:#1}}}
\def\mn@eprint@dblp#1{\href {http://dblp.uni-trier.de/rec/bibtex/#1.xml}
  {dblp:#1}}
\def\mn@eprint@#1:#2:#3:#4\@nil{\def\@tempa {#1}\def\@tempb {#2}\def\@tempc
  {#3}\ifx \@tempc \@empty \let \@tempc \@tempb \let \@tempb \@tempa \fi \ifx
  \@tempb \@empty \def\@tempb {arXiv}\fi \@ifundefined
  {mn@eprint@\@tempb}{\@tempb:\@tempc}{\expandafter \expandafter \csname
  mn@eprint@\@tempb\endcsname \expandafter{\@tempc}}}

\bibitem[\protect\citeauthoryear{{Avni}}{{Avni}}{1976}]{Avni1976}
{Avni} Y.,  1976, \mn@doi [\apj] {10.1086/154870}, \href
  {https://ui.adsabs.harvard.edu/abs/1976ApJ...210..642A} {210, 642}

\bibitem[\protect\citeauthoryear{{Barmby}, {McLaughlin}, {Harris}, {Harris}  \&
  {Forbes}}{{Barmby} et~al.}{2007}]{Barmby2007}
{Barmby} P.,  {McLaughlin} D.~E.,  {Harris} W.~E.,  {Harris} G. L.~H.,
  {Forbes} D.~A.,  2007, \mn@doi [\aj] {10.1086/516777}, \href
  {https://ui.adsabs.harvard.edu/abs/2007AJ....133.2764B} {133, 2764}

\bibitem[\protect\citeauthoryear{{Bastian}}{{Bastian}}{2016}]{Bastianrev2016}
{Bastian} N.,  2016, \mn@doi [EAS Publications Series] {10.1051/eas/1680002},
  80-81, 5

\bibitem[\protect\citeauthoryear{{Bastian} \& {Goodwin}}{{Bastian} \&
  {Goodwin}}{2006}]{BastianGoodwin2006}
{Bastian} N.,  {Goodwin} S.~P.,  2006, \mn@doi [\mnras]
  {10.1111/j.1745-3933.2006.00162.x}, \href
  {https://ui.adsabs.harvard.edu/abs/2006MNRAS.369L...9B} {369, L9}

\bibitem[\protect\citeauthoryear{{Bastian}, {Gieles}, {Goodwin}, {Trancho},
  {Smith}, {Konstantopoulos}  \& {Efremov}}{{Bastian}
  et~al.}{2008}]{BastianGieles2008}
{Bastian} N.,  {Gieles} M.,  {Goodwin} S.~P.,  {Trancho} G.,  {Smith} L.~J.,
  {Konstantopoulos} I.,   {Efremov} Y.,  2008, \mn@doi [\mnras]
  {10.1111/j.1365-2966.2008.13547.x}, \href
  {https://ui.adsabs.harvard.edu/abs/2008MNRAS.389..223B} {389, 223}

\bibitem[\protect\citeauthoryear{{Bastian} et~al.,}{{Bastian}
  et~al.}{2011}]{Bastian2011}
{Bastian} N.,  et~al., 2011, \mn@doi [\mnras]
  {10.1111/j.1745-3933.2011.01103.x}, \href
  {http://adsabs.harvard.edu/abs/2011MNRAS.417L...6B} {417, L6}

\bibitem[\protect\citeauthoryear{{Baumgardt} \& {Hilker}}{{Baumgardt} \&
  {Hilker}}{2018}]{Baumgardt2018}
{Baumgardt} H.,  {Hilker} M.,  2018, \mn@doi [\mnras] {10.1093/mnras/sty1057},
  \href {https://ui.adsabs.harvard.edu/abs/2018MNRAS.478.1520B} {478, 1520}

\bibitem[\protect\citeauthoryear{{Bertin}}{{Bertin}}{2011}]{Bertinpsfex}
{Bertin} E.,  2011, in {Evans} I.~N.,  {Accomazzi} A.,  {Mink} D.~J.,   {Rots}
  A.~H.,  eds,  Astronomical Society of the Pacific Conference Series Vol. 442,
  Astronomical Data Analysis Software and Systems XX. p.~435

\bibitem[\protect\citeauthoryear{{Bertin} \& {Arnouts}}{{Bertin} \&
  {Arnouts}}{1996}]{Bertinsextractor}
{Bertin} E.,  {Arnouts} S.,  1996, \mn@doi [\aaps] {10.1051/aas:1996164}, \href
  {http://adsabs.harvard.edu/abs/1996A%26AS..117..393B} {117, 393}

\bibitem[\protect\citeauthoryear{{Binney} \& {Tremaine}}{{Binney} \&
  {Tremaine}}{1987}]{Galdynbook}
{Binney} J.,  {Tremaine} S.,  1987, {Galactic dynamics}.
Princeton University Press

\bibitem[\protect\citeauthoryear{{Chandar}, {Ford}  \& {Tsvetanov}}{{Chandar}
  et~al.}{2001}]{Chandar2001}
{Chandar} R.,  {Ford} H.~C.,   {Tsvetanov} Z.,  2001, \mn@doi [\aj]
  {10.1086/322127}, \href
  {https://ui.adsabs.harvard.edu/abs/2001AJ....122.1330C} {122, 1330}

\bibitem[\protect\citeauthoryear{{Chandar}, {Whitmore}, {Calzetti}, {Di Nino},
  {Kennicutt}, {Regan}  \& {Schinnerer}}{{Chandar} et~al.}{2011}]{Chandar2011}
{Chandar} R.,  {Whitmore} B.~C.,  {Calzetti} D.,  {Di Nino} D.,  {Kennicutt}
  R.~C.,  {Regan} M.,   {Schinnerer} E.,  2011, \mn@doi [\apj]
  {10.1088/0004-637X/727/2/88}, \href
  {http://adsabs.harvard.edu/abs/2011ApJ...727...88C} {727, 88}

\bibitem[\protect\citeauthoryear{{Davidge}}{{Davidge}}{2008}]{Davidge2008}
{Davidge} T.~J.,  2008, \mn@doi [\aj] {10.1088/0004-6256/136/6/2502}, \href
  {https://ui.adsabs.harvard.edu/abs/2008AJ....136.2502D} {136, 2502}

\bibitem[\protect\citeauthoryear{{Djorgovski}}{{Djorgovski}}{1995}]{Djorgovski1995}
{Djorgovski} S.,  1995, \mn@doi [\apjl] {10.1086/187707}, \href
  {https://ui.adsabs.harvard.edu/abs/1995ApJ...438L..29D} {438, L29}

\bibitem[\protect\citeauthoryear{{Elson}, {Fall}  \& {Freeman}}{{Elson}
  et~al.}{1987}]{Elson}
{Elson} R.~A.~W.,  {Fall} S.~M.,   {Freeman} K.~C.,  1987, \mn@doi [ApJ]
  {10.1086/165807}, \href {http://adsabs.harvard.edu/abs/1987ApJ...323...54E}
  {323, 54}

\bibitem[\protect\citeauthoryear{{Elson}, {Schade}, {Thomson}  \&
  {Mackay}}{{Elson} et~al.}{1992}]{Elson1992}
{Elson} R.~A.~W.,  {Schade} D.~J.,  {Thomson} R.~C.,   {Mackay} C.~D.,  1992,
  \mn@doi [\mnras] {10.1093/mnras/258.1.103}, \href
  {http://adsabs.harvard.edu/abs/1992MNRAS.258..103E} {258, 103}

\bibitem[\protect\citeauthoryear{{Fall} \& {Zhang}}{{Fall} \&
  {Zhang}}{2001}]{Fall&Zhang}
{Fall} S.~M.,  {Zhang} Q.,  2001, \mn@doi [\apj] {10.1086/323358}, \href
  {http://adsabs.harvard.edu/abs/2001ApJ...561..751F} {561, 751}

\bibitem[\protect\citeauthoryear{{Forbes} et~al.,}{{Forbes}
  et~al.}{2018}]{ForbesGCreview2018}
{Forbes} D.~A.,  et~al., 2018, \mn@doi [Proceedings of the Royal Society of
  London Series A] {10.1098/rspa.2017.0616}, \href
  {https://ui.adsabs.harvard.edu/\#abs/2018RSPSA.47470616F} {474, 20170616}

\bibitem[\protect\citeauthoryear{{Freedman} et~al.,}{{Freedman}
  et~al.}{1994}]{Freedman}
{Freedman} W.~L.,  et~al., 1994, \mn@doi [\apj] {10.1086/174172}, \href
  {http://adsabs.harvard.edu/abs/1994ApJ...427..628F} {427, 628}

\bibitem[\protect\citeauthoryear{{Fujii}, {Saitoh}  \& {Portegies
  Zwart}}{{Fujii} et~al.}{2012}]{Fujii2012}
{Fujii} M.~S.,  {Saitoh} T.~R.,   {Portegies Zwart} S.~F.,  2012, \mn@doi
  [\apj] {10.1088/0004-637X/753/1/85}, \href
  {https://ui.adsabs.harvard.edu/abs/2012ApJ...753...85F} {753, 85}

\bibitem[\protect\citeauthoryear{{Gieles}}{{Gieles}}{2013}]{Gieles2013}
{Gieles} M.,  2013, Memorie della Societa Astronomica Italiana, \href
  {https://ui.adsabs.harvard.edu/abs/2013MmSAI..84..148G} {84, 148}

\bibitem[\protect\citeauthoryear{Gieles, Zwart, Baumgardt, Athanassoula,
  Lamers, Sipior  \& Leenaarts}{Gieles et~al.}{2006}]{Gieles2006}
Gieles M.,  Zwart S. F.~P.,  Baumgardt H.,  Athanassoula E.,  Lamers H. J. G.
  L.~M.,  Sipior M.,   Leenaarts J.,  2006, \mn@doi [\mnras]
  {10.1111/j.1365-2966.2006.10711.x}, 804, 793

\bibitem[\protect\citeauthoryear{{Jedrzejewski}}{{Jedrzejewski}}{1987}]{Ellipse_iraf1987}
{Jedrzejewski} R.~I.,  1987, \mn@doi [\mnras] {10.1093/mnras/226.4.747}, \href
  {https://ui.adsabs.harvard.edu/abs/1987MNRAS.226..747J} {226, 747}

\bibitem[\protect\citeauthoryear{{King}}{{King}}{1962}]{King_emp}
{King} I.,  1962, \mn@doi [AJ] {10.1086/108756}, \href
  {http://adsabs.harvard.edu/abs/1962AJ.....67..471K} {67, 471}

\bibitem[\protect\citeauthoryear{{King}}{{King}}{1966}]{King_dyn}
{King} I.~R.,  1966, \mn@doi [\aj] {10.1086/109857}, \href
  {http://adsabs.harvard.edu/abs/1966AJ.....71...64K} {71, 64}

\bibitem[\protect\citeauthoryear{{Konstantopoulos}, {Bastian}, {Smith},
  {Westmoquette}, {Trancho}  \& {Gallagher}}{{Konstantopoulos}
  et~al.}{2009}]{Konst2009}
{Konstantopoulos} I.~S.,  {Bastian} N.,  {Smith} L.~J.,  {Westmoquette} M.~S.,
  {Trancho} G.,   {Gallagher} III J.~S.,  2009, \mn@doi [\apj]
  {10.1088/0004-637X/701/2/1015}, \href
  {http://adsabs.harvard.edu/abs/2009ApJ...701.1015K} {701, 1015}

\bibitem[\protect\citeauthoryear{{Kormendy}}{{Kormendy}}{1985}]{Kormendy1985}
{Kormendy} J.,  1985, \mn@doi [\apj] {10.1086/163350}, \href
  {https://ui.adsabs.harvard.edu/abs/1985ApJ...295...73K} {295, 73}

\bibitem[\protect\citeauthoryear{{Larsen}}{{Larsen}}{1999}]{Larsenishape}
{Larsen} S.~S.,  1999, \mn@doi [\aaps] {10.1051/aas:1999509}, \href
  {http://adsabs.harvard.edu/abs/1999A%26AS..139..393L} {139, 393}

\bibitem[\protect\citeauthoryear{Lynden-Bell, Wood  \& Royal}{Lynden-Bell
  et~al.}{1968}]{Lynden-Bell1968}
Lynden-Bell D.,  Wood R.,   Royal A.,  1968, \mn@doi [\mnras]
  {10.1093/mnras/138.4.495}, 138, 495

\bibitem[\protect\citeauthoryear{{Mackey} \& {Gilmore}}{{Mackey} \&
  {Gilmore}}{2003a}]{MackeyGilmore2003a}
{Mackey} A.~D.,  {Gilmore} G.~F.,  2003a, \mn@doi [\mnras]
  {10.1046/j.1365-8711.2003.06021.x}, \href
  {https://ui.adsabs.harvard.edu/abs/2003MNRAS.338...85M} {338, 85}

\bibitem[\protect\citeauthoryear{{Mackey} \& {Gilmore}}{{Mackey} \&
  {Gilmore}}{2003b}]{MackeyGilmore2003b}
{Mackey} A.~D.,  {Gilmore} G.~F.,  2003b, \mn@doi [\mnras]
  {10.1046/j.1365-8711.2003.06022.x}, \href
  {https://ui.adsabs.harvard.edu/abs/2003MNRAS.338..120M} {338, 120}

\bibitem[\protect\citeauthoryear{{Mackey}, {Wilkinson}, {Davies}  \&
  {Gilmore}}{{Mackey} et~al.}{2008}]{Mackey2008}
{Mackey} A.~D.,  {Wilkinson} M.~I.,  {Davies} M.~B.,   {Gilmore} G.~F.,  2008,
  \mn@doi [\mnras] {10.1111/j.1365-2966.2008.13052.x}, \href
  {https://ui.adsabs.harvard.edu/abs/2008MNRAS.386...65M} {386, 65}

\bibitem[\protect\citeauthoryear{{Mayya}, {Carrasco}  \& {Luna}}{{Mayya}
  et~al.}{2005}]{Mayya2005}
{Mayya} Y.~D.,  {Carrasco} L.,   {Luna} A.,  2005, \mn@doi [\apjl]
  {10.1086/432644}, \href {http://adsabs.harvard.edu/abs/2005ApJ...628L..33M}
  {628, L33}

\bibitem[\protect\citeauthoryear{{Mayya}, {Bressan}, {Carrasco}  \&
  {Hernand\'ez-Martinez}}{{Mayya} et~al.}{2006}]{Mayya2006}
{Mayya} Y.~D.,  {Bressan} A.,  {Carrasco} L.,   {Hernand\'ez-Martinez} L.,
  2006, \mn@doi [\apj] {10.1086/506270}, \href
  {https://ui.adsabs.harvard.edu/abs/2006ApJ...649..172M} {649, 172}

\bibitem[\protect\citeauthoryear{Mayya, Romano, Rodr{\'{\i}}guez-Merino,
  {Luna}, Carrasco  \& Rosa-Gonz{\'{a}}lez}{Mayya et~al.}{2008}]{Mayyacat}
Mayya Y.~D.,  Romano R.,  Rodr{\'{\i}}guez-Merino L.~H.,  {Luna} A.,  Carrasco
  L.,   Rosa-Gonz{\'{a}}lez D.,  2008, \apj, 679, 404

\bibitem[\protect\citeauthoryear{{McLaughlin}}{{McLaughlin}}{2000}]{McLaughlin2000}
{McLaughlin} D.~E.,  2000, \mn@doi [\apj] {10.1086/309247}, \href
  {http://adsabs.harvard.edu/abs/2000ApJ...539..618M} {539, 618}

\bibitem[\protect\citeauthoryear{{McLaughlin} \& {van der Marel}}{{McLaughlin}
  \& {van der Marel}}{2005}]{McLaughlin2005}
{McLaughlin} D.~E.,  {van der Marel} R.~P.,  2005, \mn@doi [\apjs]
  {10.1086/497429}, \href {http://adsabs.harvard.edu/abs/2005ApJS..161..304M}
  {161, 304}

\bibitem[\protect\citeauthoryear{{McLaughlin}, {Barmby}, {Harris}, {Forbes}  \&
  {Harris}}{{McLaughlin} et~al.}{2008}]{Mclaughlin2008}
{McLaughlin} D.~E.,  {Barmby} P.,  {Harris} W.~E.,  {Forbes} D.~A.,   {Harris}
  G.~L.~H.,  2008, \mn@doi [\mnras] {10.1111/j.1365-2966.2007.12566.x}, \href
  {http://adsabs.harvard.edu/abs/2008MNRAS.384..563M} {384, 563}

\bibitem[\protect\citeauthoryear{{Melo}, {Mu{\~n}oz-Tu{\~n}{\'o}n},
  {Ma{\'{\i}}z-Apell{\'a}niz}  \& {Tenorio-Tagle}}{{Melo}
  et~al.}{2005}]{Melo2005}
{Melo} V.~P.,  {Mu{\~n}oz-Tu{\~n}{\'o}n} C.,  {Ma{\'{\i}}z-Apell{\'a}niz} J.,
  {Tenorio-Tagle} G.,  2005, \mn@doi [\apj] {10.1086/426421}, \href
  {http://adsabs.harvard.edu/abs/2005ApJ...619..270M} {619, 270}

\bibitem[\protect\citeauthoryear{{Moreno}, {Pichardo}  \&
  {Vel{\'a}zquez}}{{Moreno} et~al.}{2014}]{MorenoPichardo2014}
{Moreno} E.,  {Pichardo} B.,   {Vel{\'a}zquez} H.,  2014, \mn@doi [The
  Astrophysical Journal] {10.1088/0004-637X/793/2/110}, \href
  {https://ui.adsabs.harvard.edu/abs/2014ApJ...793..110M} {793, 110}

\bibitem[\protect\citeauthoryear{{Mutchler} et~al.,}{{Mutchler}
  et~al.}{2007}]{Mutchler2007}
{Mutchler} M.,  et~al., 2007, \mn@doi [\pasp] {10.1086/511160}, \href
  {https://ui.adsabs.harvard.edu/abs/2007PASP..119....1M} {119, 1}

\bibitem[\protect\citeauthoryear{{O'Connell}, {Gallagher}, {Hunter}  \&
  {Colley}}{{O'Connell} et~al.}{1995}]{Oconnell1995}
{O'Connell} R.~W.,  {Gallagher} John~S. I.,  {Hunter} D.~A.,   {Colley} W.~N.,
  1995, \mn@doi [\apjl] {10.1086/187916}, \href
  {https://ui.adsabs.harvard.edu/abs/1995ApJ...446L...1O} {446, L1}

\bibitem[\protect\citeauthoryear{{Parker}, {Wright}, {Goodwin}  \&
  {Meyer}}{{Parker} et~al.}{2014}]{Parker2014}
{Parker} R.~J.,  {Wright} N.~J.,  {Goodwin} S.~P.,   {Meyer} M.~R.,  2014,
  \mn@doi [\mnras] {10.1093/mnras/stt2231}, \href
  {https://ui.adsabs.harvard.edu/abs/2014MNRAS.438..620P} {438, 620}

\bibitem[\protect\citeauthoryear{{Peng}, {Ho}, {Impey}  \& {Rix}}{{Peng}
  et~al.}{2010}]{Penggalfit}
{Peng} C.~Y.,  {Ho} L.~C.,  {Impey} C.~D.,   {Rix} H.-W.,  2010, \mn@doi [\aj]
  {10.1088/0004-6256/139/6/2097}, \href
  {http://adsabs.harvard.edu/abs/2010AJ....139.2097P} {139, 2097}

\bibitem[\protect\citeauthoryear{{Portegies Zwart}, {McMillan}  \&
  {Gieles}}{{Portegies Zwart} et~al.}{2010}]{Portegiesrev}
{Portegies Zwart} S.~F.,  {McMillan} S.~L.~W.,   {Gieles} M.,  2010, \mn@doi
  [\araa] {10.1146/annurev-astro-081309-130834}, \href
  {http://adsabs.harvard.edu/abs/2010ARA%26A..48..431P} {48, 431}

\bibitem[\protect\citeauthoryear{Press, Teukolsky, Vetterling  \&
  Flannery}{Press et~al.}{1992}]{Numrec}
Press W.,  Teukolsky S.,  Vetterling W.,   Flannery B.,  1992, Numerical
  Recipes in Fortran 77: The art of Scientific Computing, 2 edn.
Cambridge University Press

\bibitem[\protect\citeauthoryear{Ryon et~al.,}{Ryon et~al.}{2015}]{Ryon}
Ryon J.~E.,  et~al., 2015, \mn@doi [MNRAS] {10.1093/mnras/stv1282}, \href
  {http://adsabs.harvard.edu/abs/2015MNRAS.452..525R} {452, 525}

\bibitem[\protect\citeauthoryear{{Ryon} et~al.,}{{Ryon}
  et~al.}{2017}]{Ryon2017}
{Ryon} J.~E.,  et~al., 2017, \mn@doi [\apj] {10.3847/1538-4357/aa719e}, \href
  {https://ui.adsabs.harvard.edu/abs/2017ApJ...841...92R} {841, 92}

\bibitem[\protect\citeauthoryear{{Santiago-Cort{\'e}s}, {Mayya}  \&
  {Rosa-Gonz{\'a}lez}}{{Santiago-Cort{\'e}s} et~al.}{2010}]{SantiagoCortes2010}
{Santiago-Cort{\'e}s} M.,  {Mayya} Y.~D.,   {Rosa-Gonz{\'a}lez} D.,  2010,
  \mn@doi [\mnras] {10.1111/j.1365-2966.2010.16531.x}, \href
  {http://adsabs.harvard.edu/abs/2010MNRAS.405.1293S} {405, 1293}

\bibitem[\protect\citeauthoryear{{Silich} \& {Tenorio-Tagle}}{{Silich} \&
  {Tenorio-Tagle}}{2017}]{Silich2017}
{Silich} S.,  {Tenorio-Tagle} G.,  2017, \mn@doi [\mnras]
  {10.1093/mnras/stw2879}, \href
  {https://ui.adsabs.harvard.edu/abs/2017MNRAS.465.1375S} {465, 1375}

\bibitem[\protect\citeauthoryear{{Sirianni} et~al.,}{{Sirianni}
  et~al.}{2005}]{Sirianni2005}
{Sirianni} M.,  et~al., 2005, \mn@doi [\pasp] {10.1086/444553}, \href
  {http://adsabs.harvard.edu/abs/2005PASP..117.1049S} {117, 1049}

\bibitem[\protect\citeauthoryear{{Smith} \& {Gallagher}}{{Smith} \&
  {Gallagher}}{2001}]{SmithGallagher2001}
{Smith} L.~J.,  {Gallagher} J.~S.,  2001, \mn@doi [\mnras]
  {10.1046/j.1365-8711.2001.04627.x}, \href
  {https://ui.adsabs.harvard.edu/abs/2001MNRAS.326.1027S} {326, 1027}

\bibitem[\protect\citeauthoryear{{Sollima}, {Baumgardt}, {Zocchi}, {Balbinot},
  {Gieles}, {H{\'e}nault-Brunet}  \& {Varri}}{{Sollima}
  et~al.}{2015}]{Sollima2015}
{Sollima} A.,  {Baumgardt} H.,  {Zocchi} A.,  {Balbinot} E.,  {Gieles} M.,
  {H{\'e}nault-Brunet} V.,   {Varri} A.~L.,  2015, \mn@doi [\mnras]
  {10.1093/mnras/stv1079}, \href
  {https://ui.adsabs.harvard.edu/abs/2015MNRAS.451.2185S} {451, 2185}

\bibitem[\protect\citeauthoryear{Spitzer}{Spitzer}{1987}]{Spitzer_book_gcs}
Spitzer L.~S.,  1987, Dynamical Evolution of Globular Clusters.
Princeton University Press

\bibitem[\protect\citeauthoryear{{Walker}, {Longmore}, {Bastian}, {Kruijssen},
  {Rathborne}, {Jackson}, {Foster}  \& {Contreras}}{{Walker}
  et~al.}{2015}]{Walker2015}
{Walker} D.~L.,  {Longmore} S.~N.,  {Bastian} N.,  {Kruijssen} J.~M.~D.,
  {Rathborne} J.~M.,  {Jackson} J.~M.,  {Foster} J.~B.,   {Contreras} Y.,
  2015, \mn@doi [\mnras] {10.1093/mnras/stv300}, \href
  {https://ui.adsabs.harvard.edu/abs/2015MNRAS.449..715W} {449, 715}

\bibitem[\protect\citeauthoryear{Wall \& Jenkins}{Wall \&
  Jenkins}{2003}]{Wall_statistics}
Wall J.~V.,  Jenkins C.~R.,  2003, {Practical statistics for astronomers}.
Cambridge Observing Handbooks for Research Astronomers, Cambridge Univ. Press,
  Leiden

\bibitem[\protect\citeauthoryear{Wang \& Ma}{Wang \& Ma}{2013}]{Wang_2013}
Wang S.,  Ma J.,  2013, \mn@doi [The Astronomical Journal]
  {10.1088/0004-6256/146/2/20}, 146, 20

\bibitem[\protect\citeauthoryear{{Whitmore} \& {Schweizer}}{{Whitmore} \&
  {Schweizer}}{1995}]{Whitmore&Schweizer1995}
{Whitmore} B.~C.,  {Schweizer} F.,  1995, \mn@doi [\aj] {10.1086/117334}, \href
  {http://adsabs.harvard.edu/abs/1995AJ....109..960W} {109, 960}

\bibitem[\protect\citeauthoryear{{Whitmore} et~al.,}{{Whitmore}
  et~al.}{2016}]{Whitmore2016}
{Whitmore} B.~C.,  et~al., 2016, \mn@doi [\aj] {10.3847/0004-6256/151/6/134},
  \href {https://ui.adsabs.harvard.edu/abs/2016AJ....151..134W} {151, 134}

\bibitem[\protect\citeauthoryear{{Wilson}}{{Wilson}}{1975}]{Wilson_dyn}
{Wilson} C.~P.,  1975, \mn@doi [\aj] {10.1086/111729}, \href
  {http://adsabs.harvard.edu/abs/1975AJ.....80..175W} {80, 175}

\bibitem[\protect\citeauthoryear{{Yun}}{{Yun}}{1999}]{Yun1999}
{Yun} M.~S.,  1999, in {Barnes} J.~E.,  {Sanders} D.~B.,  eds,  IAU Symposium
  Vol. 186, Galaxy Interactions at Low and High Redshift. p.~81

\bibitem[\protect\citeauthoryear{{de Vaucouleurs}, {de Vaucouleurs}, {Corwin},
  {Buta}, {Paturel}  \& {Fouque}}{{de Vaucouleurs} et~al.}{1991}]{rc31991}
{de Vaucouleurs} G.,  {de Vaucouleurs} A.,  {Corwin} Herold~G. J.,  {Buta}
  R.~J.,  {Paturel} G.,   {Fouque} P.,  1991, {Third Reference Catalogue of
  Bright Galaxies}

\makeatother
\end{thebibliography}

\appendix

\section{Ellipticity distribution of M82 disk SSCs}

\begin{figure}
\begin{center}
\includegraphics[width=\columnwidth]{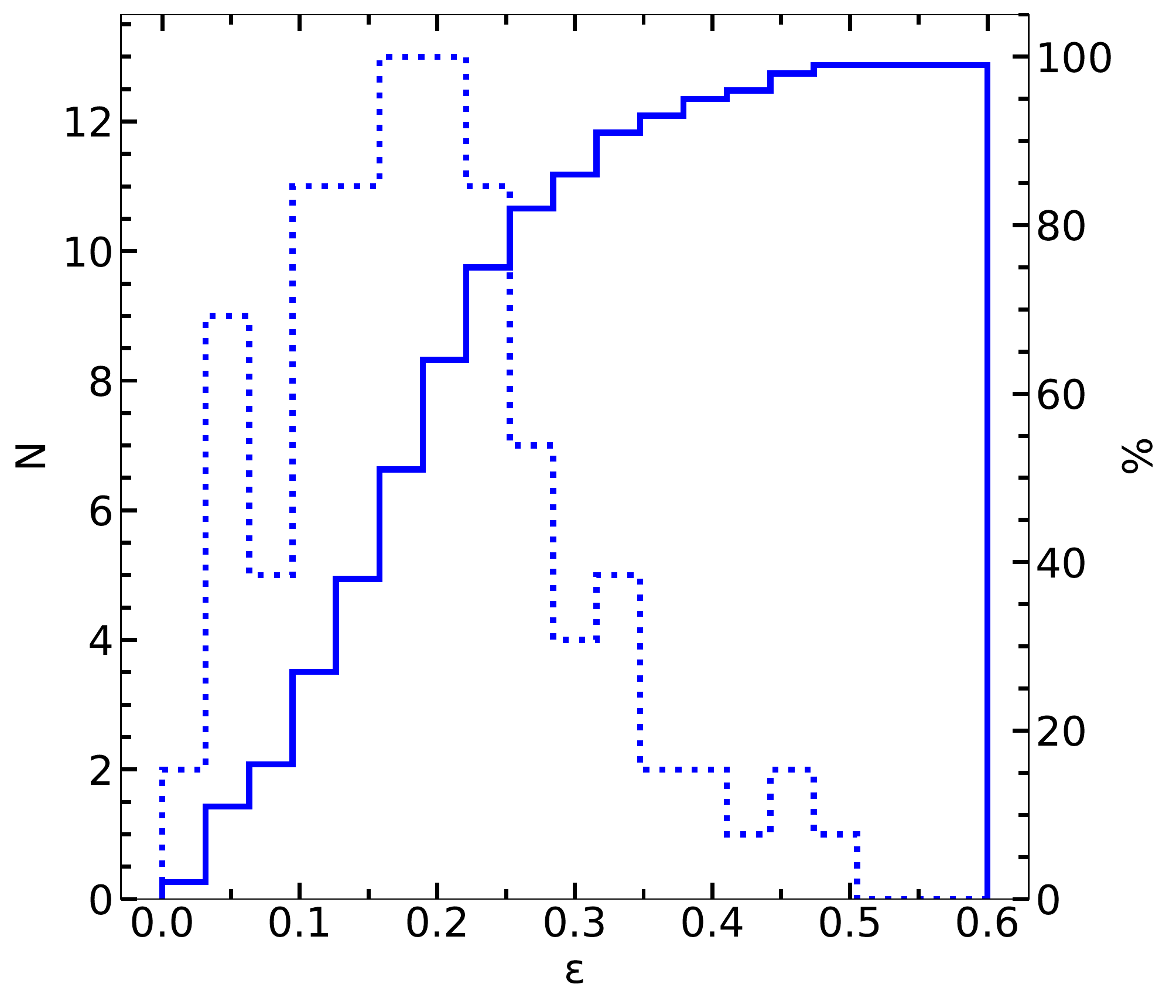}
\caption{Distribution of ellipticities measured using elliptical isophotes of the sample of 99 M82 disk SSCs (dotted line).  The cumulative distribution is shown with a solid line.}
\label{fig:profellip}
\end{center}
\end{figure}

Observed structures of stellar clusters are best described by isothermal models which have intrinsically axially symmetric radial intensity profiles \citep[e.g.][]{King_dyn}.  However, observed clusters are not always spherically symmetric.  In such cases, it is a common practice to obtain radial intensity profiles of observed clusters using circularly symmetric isophotes.  In this appendix, we discuss the effect of obtaining surface brightness profiles (SBPs) using almost circular isophotes for clusters that may have a non-negligible ellipticity.  In Fig. \ref{fig:profellip}, we show the distribution of ellipticity for our sample of 99 SSCs in the disk of M82.  These ellipticities are measured at the semi-major axis=\Rfit\ value for each cluster in the V-band using the IRAF/STSDAS task {\it ellipse}.  The plotted value corresponds to the average of ellipticities at three consecutive ellipses centered at \Rfit.  In the figure, we also show the cumulative distribution of ellipticity. The distribution of ellipticities peaks at 0.19  with only 25\% of the 
SSCs having higher ellipticities.  Thus, the majority of the SSCs are nearly circular.  

\begin{figure}
\begin{center}
\includegraphics[width=.45\textwidth]{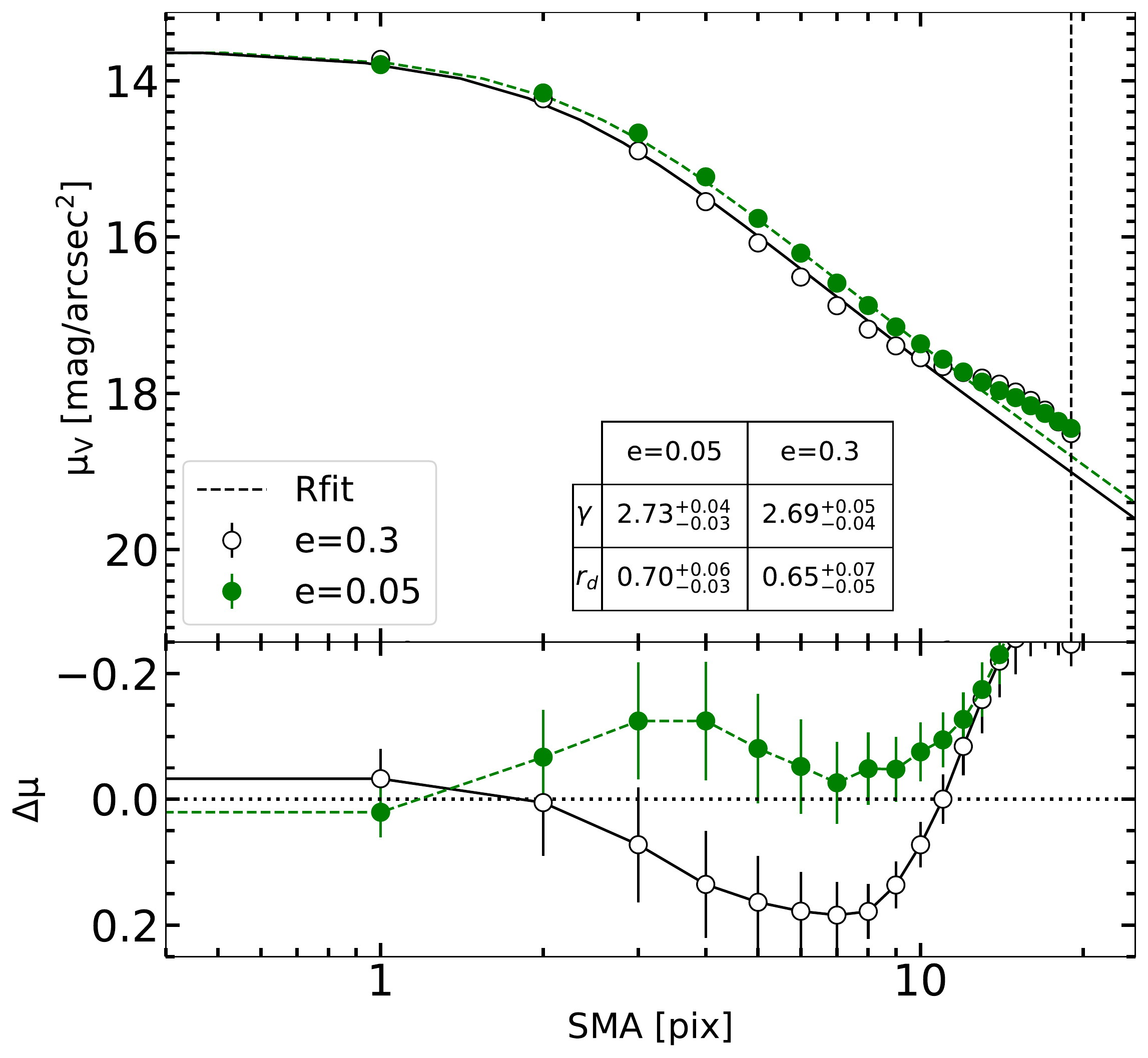}
\caption{Surface brightness profiles for the cluster D1, with nearly circular  (e=0.05) and with elliptical (e=0.3) isophotes (upper panel) and the corresponding residuals (bottom panel).}
\label{fig:profellip2}
\end{center}
\end{figure}

In Figure \ref{fig:profellip2}, we illustrate the effect of using almost circular rings ($\epsilon$=0.05) for measuring the SBP of a cluster that has $\epsilon=0.30$.   We chose well-known SSC M82-F (D1), one of the most elongated clusters, for illustration. We follow the same procedure as explained in Sec. \S\ref{Sec:method_fit} to fit the profiles using Moffat-EFF model to the SBP obtained from $\epsilon=0.30$.  The observed SBPs obtained with $\epsilon=0.05$ and $\epsilon=0.3$, along with the respective best-fit models are shown in the figure.  The $r_{\rm d}$ and $\gamma$ values for these two SBPs are identical within the errors of the measurements. This illustrates that the derived structural parameters are not very sensitive to small differences in ellipticities.
Hence, obtaining SBPs using circular apertures gives equally good values for clusters with $\epsilon$ as large as $\sim$0.3.

\begin{table}
\begin{center}
\caption{Geometrical properties of M82 disk SSCs.}
\label{tab:geo_pars}
\begin{tabular}{lrr}
\hline
 ID & $\epsilon$ & P.A.\\
(1) & (2) & (3)\\ 
\hline
D1    &   0.30  & 56\\
D4    &    0.16  & 5\\
D7      &  0.41 &   44\\
D8      &  0.20  &  58\\
D10     &  0.24 & 53\\
D14    &   0.23 & $-$25\\
D15     &   0.40 &  71\\
\hline
\end{tabular}
\end{center}
\hfill\parbox[t]{\columnwidth}{Col (1): Numerical ID, taken from \citet{Mayyacat}.  Col (2): Measured ellipticity.  Col (3): Measured position angle in degrees. The full table is shown in the electronic edition;  a portion is shown here for guidance.}
 \end{table}

\end{document}